\documentclass[twocolumn]{article}
\usepackage{graphicx}
\usepackage{amssymb}
\usepackage{times}

\newcommand{\AHEP}{AHEP Group, Instituto de F\'{\i}sica Corpuscular,
  C.S.I.C./Universitat de Val{\`e}ncia \\
\small\it  Edificio de Institutos de Paterna, Apartado 22085,
  E--46071 Val{\`e}ncia, Spain}

\def\sgn{{\rm sgn}}

\renewcommand{\Re}{\mathop{\rm Re}\nolimits}
\renewcommand{\Im}{\mathop{\rm Im}\nolimits}

\newcommand{\npb}{Nucl. Phys. B}
\newcommand{\plb}{Phys. Lett. B}
\newcommand{\prd}{Phys. Rev. D}
\newcommand{\mpla}{Mod. Phys. Lett. A}
\newcommand{\rmp}{Rev. Mod. Phys.}
\newcommand{\apj}{ApJ}
\newcommand{\aap}{A\&A}
\newcommand{\mnras}{MNRAS}
\newcommand{\lsim}{\lesssim}

\newcommand{\ga}{\gtrsim}

\def\ssubsubsection#1{\vspace{3mm} \noindent \textbf{#1} \\ \vspace{-3mm} \\ \noindent}
\def\vector#1{{\bf{#1}}}
\def\rot{{\rm rot}\,}
\def\div{{\rm div}\,}
\def\pref#1{(\ref{#1})}
\def\kb{k_{\scriptscriptstyle B}}
\def\ignore#1{}  

\title{\LARGE Resonant origin for density fluctuations deep within the Sun:
helioseismology and magneto-gravity waves}

\author{C.P.~Burgess$^{1}$\thanks{E-mail:
        cliff@physics.mcgill.ca (CPB); namig@izmiran.rssi.ru (NSD);
        timur@ific.uv.es (TIR); semikoz@ific.uv.es (VBS);
        valle@ific.uv.es (JWFV)},
        N.S.~Dzhalilov$^{2}$\footnotemark[1],
        T.I.~Rashba$^{2,3}$\footnotemark[1],
        V.B.~Semikoz$^{2,3}$\footnotemark[1],
        J.W.F.~Valle$^{3}$\footnotemark[1]\vspace{0.5cm}\\
\small\it
        $^{1}$Physics Department, McGill University, 3600 University
\small\it
        Street, Montr\'eal, Qu\'ebec, Canada, H3A 2T8\hfill\\ 
\small\it
        $^{2}$The Institute of Terrestrial Magnetism, Ionosphere and
\small\it
        Radio Wave Propagation of the Russian Academy of Sciences,\hfill\\
\small\it
        IZMIRAN, Troitsk, Moscow region, 142190, Russia\hfill\\
\small\it
        $^{3}$\AHEP\hfill}

\begin{document}
\date{}
\maketitle

{\noindent IFIC-03-12; McGill-02/40 \hfill}\\



\abstract{
  We analyze helioseismic waves near the solar equator in the presence
  of magnetic fields deep within the solar radiative zone.  We find
  that reasonable magnetic fields can significantly alter the shapes
  of the wave profiles for helioseismic $g$-modes. They can do so
  because the existence of density gradients allows $g$-modes to
  resonantly excite Alfv\'en waves, causing mode energy to be
  funnelled along magnetic field lines, away from the solar equatorial
  plane. The resulting wave forms show comparatively sharp spikes in
  the density profile at radii where these resonances take place. We
  estimate how big these waves might be in the Sun, and perform a
  first search for observable consequences.  We find the density
  excursions at the resonances to be too narrow to be ruled out by
  present-day analyses of $p$-wave helioseismic spectra, even if their
  amplitudes were to be larger than a few percent. (In contrast it has
  been shown in \cite{Burgess3} that such density excursions could
  affect solar neutrino fluxes in an important way.)  Because solar
  $p$-waves are not strongly influenced by radiative-zone magnetic
  fields, standard analyses of helioseismic data should not be
  significantly altered.  The influence of the magnetic field on the
  $g$-mode frequency spectrum could be used to probe sufficiently
  large radiative-zone magnetic fields should solar $g$-modes ever be
  definitively observed. Our results would have stronger implications
  if overstable solar $g$-modes should prove to have very large
  amplitudes, as has sometimes been argued.}

\section{Introduction}

Helioseismology has become a precision tool for studying the inner
workings of the Sun, providing one of the only direct probes of
physical properties as a function of depth, below the convective layer
and into the radiative zone. The frequencies of thousands of normal
modes have been measured, and current levels of agreement between
these measurements and calculations preclude deviations of density
profiles and sound speeds (as functions of depth) in excess of one
percent from solar-model predictions. Careful analysis of small
discrepancies between models and measurements which have arisen in the
past has helped refine the models, by winnowing out small errors in
opacities and other quantities used as inputs.

The vast majority of helioseismic analyses to date are performed in
the approximation which neglects the sun's magnetic field. This
approximation is generally very reasonable since the energy density,
$\vector{B}^2/8\pi$, of the expected fields is much smaller than other
energies in the problem, such as gas pressures, deep in the solar
interior. This is expected to be particularly true in the solar
radiative zone, below the turbulent convection which is believed to be
responsible for the solar dynamo.

Very little is directly known about magnetic field strengths within
the radiative zone -- see \cite{Cowling,Bahcall71} for early
studies. A generally-applicable bound is due to Chandrasekhar, and
states that the magnetic field energy must be less than the
gravitational binding energy: $B^2/8\pi \lsim GM^2_\odot/R_\odot^4$, or
$B \lsim 10^8$ G. Stronger bounds are possible if one makes assumptions
about the nature and origins of the solar magnetic field. For
instance, if it is a relic of the primordial field of the collapsing
gas cloud from which the sun formed \cite{Parker}, then it has been
argued that central fields cannot exceed around 30 G \cite{Boruta}.
Similarly, the dynamo mechanism can only generate a global field in
the radiative zone of a newly-born Sun with amplitude below 1 G
\cite{Kitchatinov}.  Even stronger limits, $B \lsim 10^{-6}$ G
apply~\cite{MestelWeiss} if the solar core is rapidly rotating, as is
sometimes proposed. On the other hand, it has recently been argued
\cite{Friedland} that fields up to 7~MGauss could persist in the
radiative zone for billions of years and are consistent with current
observational bounds. Other authors \cite{Couvidat} have recently
entertained radiative-zone fields as large as 30~MGauss.  Since the
initial origin and current nature of the central magnetic field is
unclear, we consider as admissible any magnetic field smaller than
of order 10~MGauss.

The purpose of the paper you are now reading is to examine the
influence of magnetic fields in more detail, deep within the solar
radiative zone. Our motivation for so doing is threefold. First,
although magnetic fields in the radiative zone are believed to be
small, they cannot be directly measured, and so quantitative bounds on
their size require the explicit calculation of their effects. Ours is
a first step along these lines, where we assume a particularly simple
geometry which we argue to be appropriate when the background magnetic
field is orthogonal to the local density gradients. Such configuration
would occur near the equatorial plane in a dipole magnetic field.
Second, plasmas in magnetic fields are notoriously unstable, with
small perturbations often giving rise to strong effects, perhaps
amplifying the implications of a small magnetic field. Indeed our
analysis uncovers such instabilities, which give rise to surprisingly
large effects. Third, the recent advent of neutrino telescopes opens a
new window on the solar interior, motivating the exploration of the
implications of solar magnetic fields for solar-neutrino properties.

In a nutshell, our main findings are these:
\begin{itemize}
\item We find, contrary to naive expectations, that profiles of
  helioseismic $g$-modes as a function of solar depth can be
  appreciably altered in the radiative zone even for magnetic fields
  which are not unreasonably large \cite{Parker,Boruta,Kitchatinov}.
  In particular, density profiles due to these waves tend to form
  comparatively narrow spikes at specific radii within the sun,
  corresponding to radii where the frequencies of magnetic Alfv\'en
  modes cross those of buoyancy-driven ($g$-type) helioseismic modes.
  Due to this resonance, energy initially in $g$-modes 
  (presumably due to turbulence at the bottom of the convective
   zone) is directly pumped into the Alfv\'en waves, causing an
  amplification of the density profiles in the vicinity of the
  resonant radius. This amplification continues until it is balanced
  by dissipation, resulting in an unexpectedly large density excursion
  at the resonant radii.
\item Since the only such level crossing which occurs within the
  radiative zone occurs with $g$-modes, and not $p$-modes, the latter
  are less substantially affected by radiative-zone magnetic fields.
  This makes it very unlikely that radiative-zone magnetic fields can
  significantly alter standard analyses of helioseismic data, which
  ignore solar magnetic fields.
\item We identify the implications of magnetic fields for the
amplitudes and frequencies of solar $g$-modes, in principle
permitting inferences about radiative-zone fields to be drawn once
these modes are experimentally observed. We find that the bound
implied by the consistency of our analysis with existing
discussions of $p$-mode helioseismic waves is at present
uninterestingly weak.
\end{itemize}

The main message in these findings is that the detailed shape of
helioseismic $g$-waves can be significantly changed by reasonable
radiative-zone magnetic fields, and this sensitivity to magnetic
fields is driven by the occurrence of level crossing between
helioseismic $g$-modes and Alfv\'en waves. Of course, although
this cartoon describes the Alfv\'en and helioseismic waves as if
they were separate, a real calculation should simply diagonalize
the normal modes of the linearized hydrodynamic problem including
the magnetic fields. We perform such a diagonalization explicitly,
and find the magneto-gravity (MG) waveforms whose peaks correspond
to the above resonance argument. This diagonalization allows us to
determine how the mode energies and excitation rates depend on the
strength of the background magnetic field.

The resonance we identify has also been recognized as possibly playing
a role elsewhere in the Sun. In particular, the feeding of energy into
this resonance has been proposed
\cite{Ionson78,Ionson82,Ionson84,Hollweg87a,Hollweg87b} as a
mechanism for heating the solar corona. Indeed the analysis of these
authors is very similar to the one we present here, with the main
difference being their neglect of gravitational forces (as is
appropriate for applications to the solar atmosphere, but not for the
radiative zone).

Indeed, we believe the study of the implications for neutrinos of the
Alfv\'en resonant layers is well worth pursuing, particularly since
density fluctuations themselves can have effects, without invoking the
existence of large magnetic moments. Two other lines of argument also
encourage such a study. First, previous experience along these lines
\cite{Krastev89,Krastev91,Schafer,Sawyer,Balantekin,Hiroshi,Burgess1}
argues that neutrinos are most sensitive to density fluctuations near
the depths at which MSW resonant oscillations take place, $r \sim 0.3
\, R_\odot$, where the resonances we derive indeed can arise. Second,
the density spikes we here find are much narrower than are generic
helioseismic $g$-wave profiles, possibly permitting Alfv\'en waves to
cause effects which garden-variety $g$-waves cannot \cite{Burgess2}.
We present a discussion of neutrino propagation through these waves in
Ref. \cite{Burgess3}, where we find potentially interesting prospects
for detectable effects if radiative-zone magnetic fields lie in the
range 10 -- 100~kG.

We organize our presentation as follows. In section 2 we set up
the hydrodynamic problem in the presence of gravity and magnetic
fields, and give the total set of MG equations whose solution is
required. We next derive the linearized MHD equations, which we
solve by relating all physical quantities to the radial component
of the perturbation magnetic field amplitude. In section 3, we
apply reasonable boundary conditions to set up an eigenvalue
problem which determines the dispersion relation for the discrete
eigenfrequency spectrum, $\omega_n(B_0)$. For a simple geometry we
identify the labels, $n$, and spectrum of the MG eigenwaves.  In
section 4 we focus on the resonance layer position and compute how
its position and width vary from mode to mode. We also here
present results for the size of the maximum density excursions at
the resonant points as a function of magnetic field strength.
Finally, section 5 discusses some observational constraints on the
wave frequencies and amplitudes.

\vskip 0.3cm
\section{The master equation for MG waves}
\vskip 0.3cm
In this section we briefly review how magnetic fields influence the
equations of hydrostatic equilibrium on which helioseimic analyses are
based. To this end we first set up the relevant MHD equations, and
linearize them to identify the differential operator whose spectrum
dictates the allowed MG wave frequencies.

Two equations of fluid dynamics are not changed by the presence of
a magnetic field -- at least within the regime of present
interest. The first of these is the continuity equation,
expressing conservation of mass:
\begin{equation}
\frac{d\rho}{dt} + \rho u = 0~ ,
\label{mass}
\end{equation}
where $d/dt = \partial/\partial t + \vector{v}\cdot \nabla$ is the
usual convective derivative, with $\vector{v}$ representing the fluid
velocity, and $\rho$ and $p$ are the fluid's mass density and
pressure. The variable $u = \div \vector{v}$ need not vanish if the
fluid is compressible~\footnote{This variable $u$ was the main focus
  of our earlier asymptotic analysis of MG waves
  \cite{DzhalilovSemikoz}.}

The second unchanged equation expresses energy conservation for a
polytropic medium. This equation states:
\begin{equation}
\frac{dp}{dt} - \gamma\frac{p}{\rho}\frac{d\rho}{dt} = -
(\gamma - 1)Q~, \label{pressure}
\end{equation}
where $\gamma = c_p/c_V$ is given by the ratio of heat capacities,
and $Q$ is the sum of all energy density sources and losses. In
principle, electromagnetic processes enter into
Eq.~\pref{pressure} through their contributions to $Q$, but for
ideal MHD we neglect both the heat conductivity and viscosity
contributions to energy losses, as well as the ohmic dissipation,
$Q = j^2/\sigma_{cond}$, where $\sigma_{cond}$ is the electrical
conductivity.

The magnetic fields enter more directly into the Euler equation
(conservation of momentum), which is of the form
\begin{equation}
\rho \frac{d\vector{v}}{dt} = - \nabla p  + \rho \vector{g} +
\frac{1}{4\pi }\Bigl [\rot \vector{B}\times \vector{B}\Bigr ]~, 
\label{Euler}
\end{equation}
with the local force of gravity given by $\rho \vector{g}$. The
local acceleration due to gravity is related to the Newtonian
potential by $\vector{g} = \nabla \phi$, with $\phi$ given by the
Poisson equation $\nabla^2 \phi = - 4\pi G\rho$. As usual, $G$ here
denotes Newton's constant. The last term of equation \pref{Euler}
expresses the contribution of the Lorentz force to the local
momentum budget.

Finally, the system is completed by Faraday's equation,
\begin{equation}
\frac{\partial \vector{B}}{\partial t} = \rot [\vector{v}\times
\vector{B}] + \frac{c^2}{4\pi \sigma_{cond}} \nabla^2 \vector{B}~,
\end{equation}
that governs the time evolution of the magnetic field. Here
$\sigma_{cond}$ represents the fluid's conductivity. In what
follows we take the plasma to be an ideal conductor, meaning we
take $\sigma_{cond}$ to be large enough to neglect the last term
in this last equation. Clearly these equations reduce to those of
standard Helioseismology Models (SHM) in the limit $B\to 0$.

\subsection{Linearized form}

Our interest is in small-amplitude oscillations about a background
configuration, and so at this point we split all variables into
background and fluctuating quantities, $A = A_0 + A'$, with the
$A'$ denoting small fluctuations about the background value $A_0$.
Since we take our background configuration to be static,
$\vector{v}_0 = 0$, we regard $\vector{v}$ to be pure fluctuation,
but suppress the prime on $\vector{v}$ in the interests of
notational simplicity.

The system of linearized equations obtained in this way is
\begin{eqnarray} &&\frac{\partial \rho'}{\partial t} +
(\vector{v} \cdot \nabla )\rho_0 + \rho_0 \; u= 0~, \nonumber\\
&&\rho_0 \frac{\partial \vector{v}}{\partial t} +  \nabla p' -
\vector{g} \rho' -  \frac{1}{4\pi}\Bigl [\rot \vector{B'}\times
\vector{B}_0\Bigr ] = 0~,\nonumber\\
&&\frac{\partial p'}{\partial t} + (\vector{v} \cdot \nabla )p_0 +
\gamma p_0 \, u = 0~,\nonumber\\
&&\frac{\partial \vector{B}'}{\partial t} =
\rot \Bigl (\vector{v}\times \vector{B}_0 \Bigr )~,  \label{linear}
\end{eqnarray}
where, as described above, $\vector{v}$, $\rho'$, $p'$ and
$\vector{B}'$ are the small Euler perturbations.

In addition to our previously-mentioned assumptions of ideal
conductance -- {i.e.} $\sigma_{cond} \to \infty$ -- and static
backgrounds -- $\vector{v}_0 = 0$, with background quantities time
independent -- these equations make three further assumptions.  First,
they adopt the Cowling approximation, which amounts to the neglect of
perturbations of the gravitational potential, ({i.e.:} $\phi' =
0$). Second, they assume the fluctuations are adiabatic, $Q'=0$, as is
satisfied with good accuracy by high frequency oscillations. Finally,
they assume the background field to satisfy $\rot \vector{B}_0 = 0$.

The backgrounds $\rho_0, p_0$ {etc.}, must themselves satisfy
the hydrodynamic equations, which for negligible (or constant)
magnetic fields reduce to the usual equations of hydrostatic
equilibrium, $\nabla p_0 = \rho_0 \vector{g}$. Finally, we suppose
there to be no energy sources or sinks in the background, $Q_0 =
0$. Under these circumstances the background density $\rho_0$ is
as given by Standard Solar Models (SSM) \cite{Bahcall88,Bahcall95}.

If we also assume $\gamma$ to be constant -- as is a good
approximation within the whole radiative zone -- we can derive from
the system of linearized equations (\ref{linear}) an evolution
equation for the velocity field $\vector{v}$:
\begin{equation}
\frac{\partial^2\vector{v}}{\partial t^2}= \nabla (\vector{g}
\cdot \vector{v}) + (\gamma - 1)\, \vector{g} \, u + c_s^2\nabla u
- v_A^2 \, \frac{\partial }{\partial t}
\left[\frac{\vector{B_0}}{B_0} \times \rot \vector{b} \right]~,
\label{linevolv}
\end{equation}
where $\vector{b} = \vector{B}'/B_0$ and $v_A = B_0/\left(4 \pi
\rho_0\right)^{1/2}$ defines the Alfv\'en speed.

\subsection{Geometry}
%
%
To proceed further we next choose the geometry of the
solution we shall seek. Our interest is in a magnetic 
field which is perpendicular to the local density
gradients and gravitational fields. Our interest is also
in the deep interior of the radiative zone, but not right
down to the solar center.

For these purposes we can take $\vector{B}_0$ to be both constant and
approximately uniform. It then suffices to work within an
approximately rectangular, rather than cylindrical, geometry. In these
circumstances it is convenient to choose a Cartesian coordinate system
whose z-axis is the `radial' direction, as defined by (but opposite in
direction to) the local gravitational acceleration, $\vector{g}$. With
this choice we take $z \in (0, R_{\odot})$, where $z=0$ represents the
solar center and $z = R_\odot$ denotes the solar surface. We are
actually mainly interested in the radiative zone, for which $z \lsim
0.7 R_\odot$.

%
Since we have in mind magnetic fields perpendicular to
physical gradients and gravitational fields, we take
$B_0$ to lie along the x-axis and the gradients to lie
along the z-axis, leading to the components
$\vector{B}_0 = (B_0,0,0)$, and $\vector{g} =
(0,0,-g(z))$.

\subsection{Isolating variables}
Returning to our original problem, we wish to solve the linear MHD
equations without resorting to the WKB approximation. Our first
goal to this end is to eliminate all variables in terms of a
single one, which we choose to be the $z$-component of the
perturbed magnetic field, $b_z(z)$. The spectrum of MHD modes is
then found from the eigenvalues of the linear ordinary
differential equation satisfied by $b_z$.

This is most easily done by transforming $\vector{v}$ and $\vector{b}$
to Fourier space in the $x$ and $y$ directions, such as in
$\vector{v} = \vector{v}(z)\exp \Bigl (-i(\omega t - k_xx -k_yy)\Bigr )$.
In this way we find that the velocity perturbations are
given by the system
\begin{eqnarray}
v_x &=& - \;\frac{ik_x(c_s^2u -
gv_z)}{\omega^2}~, \label{vx} \\
v_y &=&
\frac{1}{\omega^2}\Bigl[-ik_y(c_s^2u - gv_z) - i\omega
v_A^2 \, \rot_z~\vector{b}\Bigr] ~, \label{vy} \\
v_z &=& \frac{1}{\omega^2}\left[\frac{\partial}{\partial z}
(gv_z) + (\gamma -
1)gu - c^2_s\frac{\partial u}{\partial z} +\right.\\ 
&&+\left.i\omega v_A^2 \,\rot_y\vector{b}\right]~. \label{vz}
\end{eqnarray}

The equation for the magnetic field perturbation $\vector{b}(z)$
is then derived from the last equation of eqs.~(\ref{linear}),
\begin{equation}
i\omega \vector{b} = ik_x\vector{v} -
\frac{\vector{B}_0}{B_0} \; u~, \label{vecb}
\end{equation}
and by the definition of the compressibility, $u$:
\begin{equation}
u = ik_xv_x + ik_yv_y + \frac{\partial v_z}{\partial z}~.
\label{u}
\end{equation}

In our later applications we are most interested in the density
perturbation given in terms of the other quantities by
\begin{equation}
i\omega \frac{\rho'}{\rho_0} = -\; \frac{\omega}{k_xH}\; b_z + u~.
\label{deltarho}
\end{equation}

Substituting the expressions for $v_{x,y}$ from  eqs.~(\ref{vx})
and (\ref{u}), and finding the magnetic field component $b_y$ from
the Maxwell equation $\div \vector{b} =0$ we find expressions for
the dynamical variables in terms of $b_z$. In so doing we also use
two further approximations. We first assume the adiabatic
parameter, $\gamma$, to be constant, and we also use the
low-frequency approximation, $\omega^2\ll k_x^2c_s^2$. (This
latter assumption has the effect of filtering out the acoustic
$p$-modes from our analysis.) We find in this way:
\begin{eqnarray} &&b_x(z) =
\frac{ik_x}{k_{\perp}^2}\frac{db_z(z)}{dz}~,\nonumber\\ &&
b_y(z) = \frac{ik_y}{k_{\perp}^2}\frac{db_z(z)}{dz}~,\nonumber\\
&&v_z(z) = \frac{\omega}{k_x}b_z(z)~,\nonumber\\
&&v_x(z) = \frac{ik_x\omega}{k_{\perp}^2}\left[ - \frac{gb_z(z)}{k_xc_s^2} +
\left(1 + \frac{\omega^2}{k_{\perp}^2c_s^2} \right)
\frac{1}{k_x}\frac{db_z(z)}{dz}\right]~,\nonumber\\
&&v_y(z) = \frac{ik_y\omega}{k_{\perp}^2}
\left[ - \frac{gb_z(z)}{k_xc_s^2} + \left(1 +
\frac{\omega^2}{k_{\perp}^2c_s^2} \right)\frac{1}{k_x}
\frac{db_z(z)}{dz}\right]~,\nonumber\\
&&\frac{p'(z)}{p_0} = i\frac{\gamma
\omega^2}{k_{\perp}^2c_s^2k_x}\frac{db_z(z)}{dz}~,\nonumber\\
&&\frac{\rho'(z)}{\rho_0} = \frac{(+i)}{k_xH}\left( \frac{\gamma -
1}{\gamma}b_z(z) +
\frac{\omega^2H}{k_{\perp}^2c_s^2}\frac{db_z(z)}{dz}\right)~,
\label{all}
\end{eqnarray}
where $k_\perp^2 = k_x^2 + k_y^2$.
Using these, expression (\ref{u}) for $u$ becomes
\begin{equation}
u(z) = \frac{\omega^3}{k_xc_s^2}\left(\frac{g}{\omega^2}b_z(z) -
\frac{1}{k_{\perp}^2}\frac{db_z(z)}{dz} \right)\neq 0 ~. \label{u1}
\end{equation}
Note that we retain here in the transverse velocity components,
$v_{x,y}$, terms which are subdominant in the quantity $
\omega^2/k_{\perp}^2c_s^2\ll 1$, since these are needed to obey
the compressibility condition, Eq.~(\ref{u}) or (\ref{u1}).

\section{The eigenvalue problem}
We may now identify the equation satisfied by $b_z(z)$ itself,
whose solution determines all of the other variables through the
expressions derived above. We find $b_z$ to be determined as the
solution to the following second order linear ordinary
differential equation,
\begin{eqnarray}
&&\left(1 - \frac{k_x^2v_A^2}{\omega^2}\right)\frac{d^2b_z(z)}{dz^2} -
\frac{N^2}{g}\frac{db_z(z)}{dz} +\nonumber\\
&&+k_{\perp}^2\left(
\frac{k_x^2v_A^2}{\omega^2} - 1 + \frac{N^2}{\omega^2} \right)b_z(z) = 0~,
\label{masterbz}
\end{eqnarray}
where $N$ denotes the Brunt-V\"ais\"al\"a frequency, as defined by
\begin{equation}
N^2(z)= g(z)\left(\frac{1}{\gamma \, p_0} \, \frac{dp_0}{dz} -
\frac{1}{\rho_0} \, \frac{d\rho_0}{dz} \right) .
\end{equation}
Equation~\pref{masterbz} is one of our main results, and the
remainder of the paper is devoted to the construction of its
solutions.

\subsection{Qualitative Analysis}
Eq.~(\ref{masterbz}) describes the interaction of adiabatic
gravity modes with magnetic fields (magneto-gravity modes), in the
approximation of low frequencies, $\omega^2\ll k_x^2c_s^2$. Before
pressing on, it is instructive to examine the qualitative
properties of the solutions to Eq.~(\ref{masterbz}), since these
capture the results we will obtain from a more detailed analysis.

\begin{figure*}
\includegraphics[width=\textwidth]{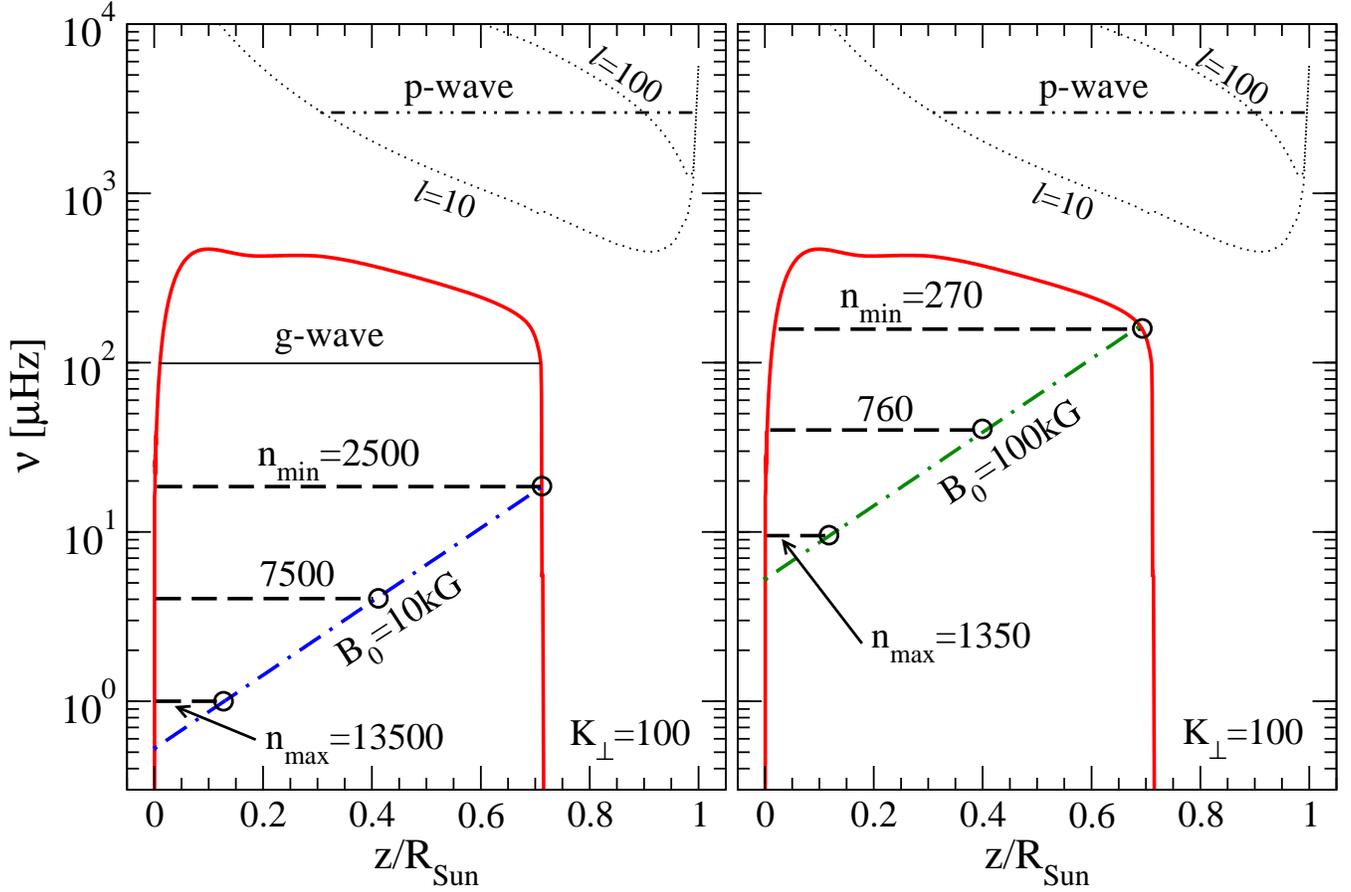} 
\caption{Relevant frequencies  plotted against radial position within
  the sun. The solid curve gives the Brunt-V\"ais\"al\"a frequency,
  $N(z)$, while the dot-dashed lines are the Alfv\'en frequencies.
  Notice that $N(z)$ goes to zero at the solar center and the top of
  the radiative zone. The horizontal dashed lines represent the
  frequencies of magneto-gravity waves computed from Eq.~(\ref{chieq})
  for several choices for the mode number, $n$, for fixed values $K_x
  = K_\perp = 100$. As discussed in the text, resonances occur where
  the mode frequencies intersect the Alfv\'en frequency, indicated by
  circles in the figure. The two panels correspond to two choices for
  the magnetic field: 10~kG (left) and 100~kG (right). Minimum and
  maximum mode numbers are indicated, with $n_{\rm min}$ defined by
  the condition that its resonance occurs at the top of the radiative
  zone, $z_r(n_{\rm min})=0.7R_\odot$, and $n_{\rm max}$ having
  resonance at $z_r(n_{\rm max})=0.12R_\odot$ (see
  Figs.~\ref{resposfig} and \ref{resspacefig}). The dotted lines
  denote acoustic (Lamb) frequencies for $l=10$ and $100$ (see
  \cite{Christensen-Dalsgaard:2002ur}) while the double-dot-dashed
  horizontal line represents the trapping region for a p-wave of
  frequency 3000~$\mu$Hz and $l=10$. The horizontal solid line
  represents the trapping region for a g-wave of frequency 
  100~$\mu$Hz.
\label{bruntfig1}}
\end{figure*}

In the limit $B_0 \to 0$ (and so $v_A \to 0$), Eq.~(\ref{masterbz})
reduces to the standard evolution equation for `pure' helioseismic
$g$-modes, which is usually expressed in terms of the variable $v_z$
(which is related to $b_z$ through the relation $v_z=\omega b_z/k_x$).
Since more detailed analyses -- see Fig.~\ref{bruntfig1} -- show
that $N$ rises from zero at the solar center, remains approximately
constant $N \approx N_0$, through the radiative zone, and then falls
to zero again at the bottom of the convective zone, these $g$-modes
can be thought of as the eigen-modes of oscillations inside the cavity
in between the two regions where $N$ goes through zero. For a given
wave frequency, $\omega < N_0$, the turning points of this cavity are
given by the condition $\omega \approx N$. For smaller $\omega$ the
lower turning point gets closer to the solar center, $z\to 0$, and the
upper one gets slightly closer to the bottom of the convective zone
(CZ).

Conversely, if gravity is turned off ($N \to 0$), then
Eq.~(\ref{masterbz}) describes Alfv\'en waves, which oscillate
with frequency $\omega = k_x v_A$ and propagate along the magnetic
field lines. Notice that since $v_A \propto \rho^{-1/2}$ this
frequency grows with $z$, since the density of the medium falls.

Keeping both magnetic and gravitational fields introduces
qualitatively new behavior, as may be seen mathematically because
equation (\ref{masterbz}) acquires a new singular point which
occurs when the coefficient of the second-derivative term
vanishes. Since this singularity appears at the Alfv\'en
frequency, 

\begin{equation}
  \label{eq:reson1}
\omega =k_xv_A  
\end{equation}
it can be interpreted as being due to a resonance between the
$g$-modes and Alfv\'en waves. This resonance turns out to occur at a
particular radius because the Alfv\'en frequency varies with radius
while the $g$-mode frequency is independent of radius (see
Fig.~\ref{bruntfig1}). The resonance occurs where the growing Alfv\'en
mode frequency crosses the frequency of one of the $g$-modes, and the
resulting waveforms would be expected to vary strongly at these
points.

The resonance can occur inside the radiative zone if the Alfv\'en
frequency climbs high enough to cross a $g$-mode frequency before
reaching the top of the radiative zone. Since $v_A \propto B_0$,
whether this occurs or not depends on the central field value, $B_0$.
Our main task is to find the dependence of the spectrum of eigen-modes
on magnetic field, and to elucidate when such singular resonances can
appear within the Sun. Of particular interest for observational
purposes is to know how the density and sound speed vary near the
resonances, especially when the resonance region occurs near the
locations of resonant neutrino oscillations.

The presence of a resonance at a particular radius, $z = z_r$, shifts
the position of the upper turning points of the corresponding
magneto-gravity wave towards deeper regions of the Sun. This leads to
several new effects. First, the shortening of the cavity due to the
presence of magnetic field causes the eigen-frequencies to depend on
the magnetic field value. Second, there can be energy transfer between
$g$-modes and Alfv\'en waves, within the narrow singular resonance
layer, leading to the corresponding eigen-frequencies acquiring
imaginary parts. Third, the nearer the upper boundary of MHD cavity is
to the solar center, the stronger the $g$-modes are confined to the
solar core.

More details of these waveforms can be obtained from a WKB-type
analysis of the master Eq.~(\ref{masterbz}). Near the solar center
where $N^2\to 0$, if $(k_xv_A)^2/\omega^2\ll 1$ one obtains the
exponential solution $b_z\to e^{\pm k_{\perp}z}$. (Which
combination of these solutions appears is fixed by boundary
conditions, {e.g.} if $b_z(0) = 0$ at the solar center we have
$b_z \propto \sinh(k_\perp z)$.) One finds similar behaviour for
the solution above the singular resonant layer, $z>z_r$, up to the
top of the radiative zone.
This exponential growth happens because of the
exponential decrease of density (and so exponential increase in
$v_A$) with $z$. The requirement for complex frequencies arises
from the demand that the solution remain regular in the narrow
Alfv\'en resonance layer.

\vskip 0.3cm
\subsection{Background Configurations}

In order to acquire analytic solutions to eq.~\pref{masterbz}, we
must specify a background profile for $\rho_0(z)$ and $g(z)$. We model
$\rho_0(z)$ as an exponential density profile, $\rho_0(z) = \rho_c \,
e^{-z/H}$ with $H$ constant, and take $g$ to be approximately
constant. We now argue these to be reasonable approximations well
inside the radiative zone, and not too close to the solar center.

The variation of background quantities may be characterized by the
scale heights, $H_A^{-1} := |(1/A_0)(dA_0/dz)|$, for $A = \rho, p,
T$ {etc.}. 
Solar models \cite{Bahcall88,Bahcall95,Stix} show that the density
scale height is the shortest in the radiative zone, $H_g,~H_T >
H_{\rho}$, and that $H_\rho$ takes the constant value $H_\rho = H =
0.1 \, R_\odot$ to a good approximation, except near the solar center
and solar surface. On these grounds we have that the gravitational
equation, $dg/dz = - 4 \pi G \, \rho_0$, implies $g = 4 \pi G \rho_c H
\, \left(1 - e^{-z/H} \right)$, which is approximately constant for
$z$ much larger than $H$. For constant $g$ the hydrostatic equation,
$dp_0/dz = - \rho_0 \, g$, then implies $p_0 = p_c \exp(-z/H)$ which
also implies constancy of the temperature for gases obeying the ideal
gas equation state, $p_0/\rho_0 = RT_0/\mu$. Here $R$ is the gas
constant, $R = \kb/m_u$, $\mu$ is the molecular weight measured in
atomic mass units $m_u\simeq m_p$, and $\kb$ is Boltzmann's constant.
In terms of these quantities the density scale height becomes: $H =
c^2/g$ (= constant) \cite{Bahcall88,Bahcall95}, with $c =
\sqrt{RT_0/\mu}$ being the isothermal sound velocity, which is related
to the adiabatic sound velocity, $c_s$, by $c_s^2 = \gamma c^2$. All
of these quantities are constants within the approximations we are
using.

Notice that our approximations that $\gamma$, $H = 0.1 R_{\odot}$
and $g$ are all constant also imply approximate constancy for the
Brunt-V\"ais\"al\"a frequency, $N$, within the radiative zone.

\vskip 0.3cm
\subsection{Solution Using Hypergeometric Functions}
Changing variables to $\xi = k_x^2 v_A^2/\omega^2 = \xi_c \,
e^{z/H}$ we recast Eq.~\pref{masterbz} as
\begin{eqnarray}
&&(1 - \xi)\xi^2 \, \frac{d^2b_z(\xi)}{d\xi^2} +
\xi \left(1 - \xi - \frac{\gamma -1}{\gamma}\right)\frac{d
b_z(\xi)}{d\xi} + \nonumber\\
&&+K_{\perp}^2(\kappa^2 + \xi - 1) \, b_z(\xi) =0~,
\label{masterbz1} 
\end{eqnarray}
where we introduce the dimensionless parameters $K_{\perp}^2 =
k_{\perp}^2H^2$, $\kappa^2 = N^2/\omega^2$.

Since the equation \pref{masterbz1} has three regular singular
points -- at $\xi = 0,1$ and $\infty$ -- it can be put into
standard hypergeometric form,
\begin{equation}
    \xi (1 - \xi) \, \frac{d^2Y}{d\xi^2} + [c - (a + b + 1)\xi] \,
    \frac{dY}{d\xi} - ab \, Y = 0~, \label{Gauss}
\end{equation}
through a change of the dependent variable $b_z = \xi^{\sigma} \,
Y(\xi)$. This permits an explicit solution for $b_z$ in terms of
Gaussian hypergeometric functions, $F(a,b;c;\xi) =
{}_2F_1(a,b;c;\xi)$. Here the hypergeometric coefficients, $a, b$
and $c$, are given by
\begin{equation}
a = \sigma + K_{\perp}, \qquad b = \sigma - K_{\perp},
\qquad c = 2\sigma + \gamma^{-1}~,
\label{hypcoeffs}
\end{equation}
where the complex index, $\sigma$, is defined by the expression
\begin{equation}
\label{sigmadef} 
2\sigma = \frac{\gamma - 1}{\gamma} -q~,
\end{equation}
with
\[
q\equiv 1-c = \left[ \left(\frac{\gamma -
1}{\gamma} \right)^2 - 4K_{\perp}^2(\kappa^2 -1) \right]^{1/2}~.\nonumber
\]

\ssubsubsection{General Solutions}
Two linearly-independent solutions to the Hypergeometric equation,
\pref{Gauss}, are
\[ Y_1=F(a,b;c;\xi)~,~~
Y_2=\xi^{1-c}F(a-c+1,b-c+1;2-c;\xi)~ .
\]
These solutions must be analytic for all $\xi \ne 0, 1$ and $\infty$,
but they may acquire branch points at these three regular singular
points depending on the values of $a$, $b$ and $c$. In the present
case because $\Re~(c-a-b) = \gamma^{-1} > 0$, the solutions are
bounded at $\xi=1$ as long as neither $c$ nor $2-c$ is zero or a
negative integer. Moreover, the second fundamental solution, $Y_2$, is
finite at $\xi=0$ if $\Re~(1-c) = \Re~ q > 0$.

We shall find that the waves of interest have complex frequencies,
which we parameterize as $\omega=\omega_1 (1+ id)$ with $\omega_1$ and
$d$ real, with $|d|\ll1$. Complex $\omega$ implies,
from~\pref{sigmadef}, that the parameters $q$ (and so also $c$) are
also complex, $q=q_1+ iq_2$, with real and imaginary parts:
\[
q=q_1 + iq_2\equiv \sqrt{w_1+iw_2}=
\]
\begin{equation} \label{qparameter}
\frac{1}{\sqrt{2}}\left
[\sqrt{\sqrt{w_1^2 + w_2^2}+w_1}+i~\sgn(d)\sqrt{\sqrt{w_1^2 +
w_2^2}-w_1}~\right]\,,
\end{equation}
where $w_{1,2}$ are given by:
\[
w_1=\left(\frac{\gamma-1}{\gamma}\right)^2 + 4K_{\perp}^2\left[ 1
- \frac{N^2\, (1-d^2)}{\omega_1^2\, (1+d^2)^2}\right]~,
\]
\[
w_2=\frac{8d \, N^2 \, K_{\perp}^2}{\omega_1^2\, (1+d^2)^2}~ .
\]
It is clear from these equations that one always has $\Re~q = q_1>0$.

The general solution for $b_z(z)$ is therefore given by the linear
combination of the independent solutions to Eq.~(\ref{Gauss}):
\begin{eqnarray}
&&b_z(z)= D_1 \, \xi^{\sigma}F(a,b;c;\xi) + \nonumber\\
&&+D_2 \,\xi^{1-\sigma -1/\gamma} F(1+a-c,1+b-c;2-c;\xi)~,
\label{solution} 
\end{eqnarray}
where one combination of the integration constants, $D_1/D_2$ is
obtained by imposing boundary conditions at $\xi_c = \xi(z = 0)$
and $\xi_* = \xi(z=z_*)$, where $z_* \approx 0.7 R_\odot$ denotes
the top of the radiative zone.

\ssubsubsection{Resonance Conditions}
Although the solutions are bounded at the singular point $\xi =
1$, they are generally {\it not} analytic there. The position
$z_r$ determined by the condition 
\begin{equation}
  \label{eq:reson2}
\hbox{Re}\, \xi(z_r) = (k_x v_A(z_r)/\omega_1)^2 = 1  
\end{equation}
therefore defines the position of the $g$-mode/Alfv\'en resonance.
This resonance occurs within the radiative zone only if $\xi_c < 1$
and $\xi_* > 1$. In this section we ask what this requires for the
properties of the wave.

Expressed in terms of the central quantities, the Alfv\'en
resonance condition is $\Re~\xi(z_r)=1$, or $\Re~\xi_c = (k_x
v_{Ac}/\omega_1)^2 = e^{-z_r/H}$. Since $0 \le z_r \le z_* \approx
0.7 R_\odot$ this condition limits the values of $K_\perp$ for
which resonances can occur. Using the numerical values $H=7\times
10^9$~cm, $v_{Ac} = (B_0/43.4~\rm{G})$~cm~s$^{-1}$ (obtained using the
central density $\rho_c = 150$~g~cm$^{-3}$), $N=10^{-3}$~s$^{-1}$
(obtained using the mean acceleration $g=0.2$~km~s$^{-2}$ and the
ideal-gas adiabatic parameter $\gamma = 5/3$), we find the
following limits on the transverse wave number $K_{\perp} \geq
K_x=k_xH$:
\begin{equation} \label{bound}
    1 \leq \frac{7\times 10^6 \, (\omega_1/N)}{K_x (B_0/43.4~\rm{G})}
    \lsim 30~,
\end{equation}
where the lower (upper) limit corresponds to the choice $z_r = 0$
($z_r = z_* \simeq 0.7R_{\odot}$).

For frequencies, $\omega_1 \sim N$, and moderate central magnetic
fields, $B_0 \lsim 100~G$, we see that the existence of a
resonance requires the wave number $K_{\perp}$ to be large,
$K_{\perp}\gg 1$. Smaller $K_\perp$ necessarily requires either
larger magnetic fields or extremely low frequencies or both. Since
we cannot, within the scope of our approximations, reduce the
angular frequency lower than the solar angular rotation frequency
(for which the Solar rotation period is $T_{\odot}\sim 27$ days),
a reasonable lower limit for $\omega_1$ might be $\omega_1 \sim
10^{-5}$~s$^{-1}$ (corresponding to a period $2\pi/\omega_1 \sim 8$
days). Even for such small frequencies one finds from
Eq.~\pref{bound} the condition $3\times 10^4 \ga K_{\perp} \ga
10^3$, as required.

\ssubsubsection{Boundary Conditions}
At the solar center ($z=0$ and so $\xi = \xi_c$) we use the
boundary condition which would have arisen as a smoothness
condition in cylindrical coordinates:
\begin{equation} b_z(0)=
k_x v_z(0)/\omega = 0 .
\end{equation}
This requires the integration constants to satisfy
\begin{equation}
\label{ratio1} \frac{D_2}{D_1}= - \xi_c^{-q}
\frac{F(a,b;c;\xi_c)}{F(1+a-c,1+b-c;2-c;\xi_c)}~.
\end{equation}

This boundary condition is sometimes called the `total reflection'
condition because of the observation that it implies (for real
$\omega$) that the reflection coefficient at $z = 0$ equals unity:
$R = |D_2/D_1|^2 = 1$. To see why this is so notice that real
$\omega$ implies $q$ is pure real or pure imaginary. Within the
domain of the approximation $k_\perp^2 c_s^2 /\omega^2 \gg 1$ it
is pure imaginary, so we write $q \to -i\beta$ for some real
$\beta$. With this choice we have $\sigma \to (\gamma
-1)/(2\gamma) + i\beta/2$, and so
\[a \to \sigma + K_{\perp}, \qquad b \to \sigma - K_{\perp}, \quad
\hbox{and} \qquad c \to 1+i \beta~, \]
which leads to the relations:
\begin{eqnarray} \label{notation}
    1+a-c &\to& \frac{\gamma -1}{2\gamma} + K_{\perp}
    - \frac{i\beta}{2} = a^*~,\nonumber\\
    1+b-c &\to& \frac{\gamma -1}{2\gamma} - K_{\perp}  -
    \frac{i\beta}{2} = b^*~,\nonumber\\
    2-c &\to&  1-i\beta~ = c^*.
\end{eqnarray}
From these conditions follows the conclusion $R=1$.

We next turn to the boundary condition at the top of the radiative
zone, $z_* \approx 0.7 R_\odot$. Since our interest is in resonant
waves which lie deep in the radiative zone, we will assume $\xi_*
= \xi(z_*) \gg 1$ when applying this boundary condition.

The behaviour of $b_z$ for $\xi>1$ is found using
Eq.~(\ref{InvIdent}) from Appendix~A together with
Eq.~(\ref{solution}):
\begin{eqnarray}
\label{evanescent}
    \frac{b_z}{D_1}&=&\left[A_1 - (-)^c  A_3 \, \frac{D_2}{D_1}
    \right] (-)^{-a}\xi^{-K_{\perp}}\times\nonumber\\
    &&\times F\left(a,1-c+a;1+a-b;\frac{1}{\xi} \right) + \nonumber\\
    &&+ \left[A_2 - (-)^c  A_4 \, \frac{D_2}{D_1} \right]
    (-)^{-b}\xi^{+K_{\perp}}\times\nonumber\\ 
    &&\times F\left(1-c+b,b;1-a+b;\frac{1}{\xi} \right)~,
\end{eqnarray}
where
\begin{eqnarray}
\label{A}
    &&A_1=\frac{\Gamma (c)\Gamma (b-a)}{\Gamma (b)\Gamma (c-a)},~~~~~A_2=
    \frac{\Gamma (c)\Gamma (a-b)}{\Gamma (a)\Gamma (c-b)},\nonumber\\
    &&A_3=\frac{\Gamma (2 -c)\Gamma (b-a)}{\Gamma (b -c +1)\Gamma
    (1-a)},\nonumber\\
    &&A_4= \frac{\Gamma (2-c)\Gamma (a-b)}{\Gamma (a-c+1)\Gamma (1-b)}~.
\end{eqnarray}

For $\xi \gg 1$, the two terms in Eq.~(\ref{evanescent}) behave as
$\xi^{\pm K_\perp} \propto e^{\pm K_\perp \, z/H}$, and so either
fall or grow exponentially as functions of $z$. We take the
absence of the growing behaviour to be our boundary condition for
$\xi=\xi_*$. This requires vanishing of the whole factor within
brackets in the second line of Eq.~(\ref{evanescent}), thereby
providing a second condition which must be satisfied by the ratio
$D_2/D_1$:
\begin{equation} \label{ratio4}
    \frac{D_2}{D_1}= (-)^{-c} \;
    \frac{\Gamma(c)\Gamma(1-b)
    \Gamma(1+a-c)}{\Gamma(2-c)\Gamma(a)\Gamma(c-b)}~.
\end{equation}
Given this boundary condition, $b_z$ falls exponentially for
$z>z_r$, $b_z\sim \xi^{-K_{\perp}} = \xi_0^{-K_{\perp}}
e^{-K_{\perp}z/H}$, as given by the first line in
Eq.(\ref{evanescent}).

\subsection{Eigenfrequencies}
The consistency of the two conditions, eqs.~\pref{ratio4} and
\pref{ratio1}, for $D_2/D_1$ gives the eigenvalue condition which
fixes the dispersion relation for the resonant waves:
\begin{eqnarray} \label{mainspectrum}
&&    \xi_c^{-q}\frac{F(a,b;c;\xi_c)}{F(1+a-c,1+b-c;2-c;\xi_c)}=\nonumber\\
&&= -(-)^{-c} \; \frac{\Gamma(c)\Gamma(1-b)
    \Gamma(1+a-c)}{\Gamma(2-c)\Gamma(a)\Gamma(c-b)} ~,
\end{eqnarray}
from which we read off the MHD frequency, $\omega(k_x,k_y,B_0,n)$,
where the integer $n$ is the mode number for the eigenfunction of
interest. The solutions which are obtained in this way have the
important property that they are complex, $\omega = \omega_1 \, (1
+ i d)$, a feature which is also present in the absence of
gravitational loading, and which has been exploited in other
contexts in attempts to explain the nature of coronal heating
\cite{Ionson78,Ionson82,Ionson84,Hollweg87a,Hollweg87b}.

We note in passing that, in contrast to the master equation in
(\ref{masterbz}), Eq.~\pref{mainspectrum} does {\em not} reduce to the
standard $g$-mode spectrum in the limit $B_0 \to 0$.  It does not do
so because of our assumption that $N$ be approximately constant.
Although this approximation suffices for the study of MG waves, it is
too crude to capture the resonant frequencies of $g$-modes. It cannot
do so because $g$-modes are trapped inside the radiative zone
precisely because $N$ is {\em not} constant, but instead drops to zero
at the top of the radiative zone. For MG waves this dropping of $N$ is
not crucial because its role in trapping the modes is instead played
by the resonant layer at $z = z_r$.

The eigenvalue condition can be written in a more transparent way by
taking advantage of some properties of hypergeometric functions.
Recall that our interest is in the regime $K_\perp N/\omega_1 \gg 1$,
and so the parameter $q$, as defined by Eq.~\pref{sigmadef}, is mostly
imaginary, $q= -i\beta$, since $|d|\ll1$, with $\beta =
2K_{\perp}N/\omega \gg 1$. This implies that $\sigma= (\gamma
-1)/(2\gamma) + i \beta/2$ is also large inasmuch as $| \sigma| \gg
1$, and so must also be $a$ and $b$. This observation allows us to use
the Watson asymptotic form for hypergeometric functions
(\ref{watson2}) -- see Appendix~A -- which applies for large values
of $a$ and $b$.

Using this, together with liberal use of the convolution identity
$\Gamma (z)\Gamma (1-z)= \pi/\sin (\pi z)$, allows us to rewrite
the eigenvalue condition, Eq.~(\ref{mainspectrum}), as a simple
transcendental equation:
\begin{equation} \label{mainspectrum1}
    -2i\beta\ln \eta = 2i\pi n + \ln \left| \tan \frac{\pi}{\gamma}
    \right| \pm i\pi\left(\frac{1-\gamma}{\gamma}\right)~,
\end{equation}
where the upper (or lower) sign on the right-hand-side is chosen
if the imaginary part of the frequency is positive: $d>0$ (or
negative, $d < 0$). The variable $\eta$ is defined by $\eta=
\xi_c^{-1/2} - \sqrt{1/\xi_c -1}$, and depends on the magnetic
field ${\bf B}_0$ through the Alfv\'en velocity $v_{Ac}$, since
$\xi_c = (k_x v_{Ac}/ \omega_1)^2$. The integer $n=0,\pm 1,\pm
2,...$ is the mode number.

The eigenvalue equation is easier to work with if re-expressed
slightly. Solving for $\xi_c(\eta)$ we find $\xi_c^{1/2} =
2\eta/(1 + \eta^2) = k_x \, v_{Ac}/\omega$, and so
\[
\beta = - \left( \frac{4HN}{\alpha \, v_{Ac}} \right) \;
\frac{\eta }{1+\eta ^{2}}
\]
where $\alpha = k_x/k_\perp \le 1$. If we now change variables to
$\chi$ defined by $\eta =e^{-\chi }$ we find
\begin{equation} \label{chieq}
    \chi =\frac{\alpha \, v_{Ac}}{4NH}\left[ 2\pi n-i\ln \left|
    \tan \frac{\pi }{\gamma }\right| \pm \pi
    \left( \frac{1-\gamma }{\gamma }\right) \right] \cosh \chi~.
\end{equation}
The variable $\chi$ is convenient because physical quantities have
simple expressions in terms of it. In particular, the phase
velocity is $v_{ph}= \omega_1 / k_x = v_{Ac}\Re\cosh \chi$; the
frequency's imaginary part is determined by $d = \Im\cosh \chi /
\Re\cosh \chi$, and the position of the resonant layer is
$z_{r}=2H \ln \left[ \Re\cosh \chi \right]$.

Notice that Eq.~\pref{chieq} takes the form $\chi = A \cosh\chi$.  For
real $A$ this equation has two real roots for $\chi$ if $A < A_{max}$
and none if $A > A_{max}$, where $A_{max}$ is defined as the maximum
value taken by the function $\chi/\cosh\chi$ (which numerically occurs
for $\chi = \chi_{max} = 1.1997)$. A similar result holds for $A$ and
$\chi$ close to real, such as happens when $\omega = \omega_1 (1 +
id)$, and so solutions with $d$ small only exist up to a maximum mode
number, $|n| \leq n_{max}$, where
\[ n_{max} = \left( \frac{2NH}{\pi \alpha \,
v_{Ac}} \right) \frac{\chi _{max}}{\cosh \chi _{max}}~.
\]
Solutions with large $d$ must be discarded because they lie beyond
the scope of the linearized approximations within which we work.

\begin{figure}
\includegraphics[width=\columnwidth]{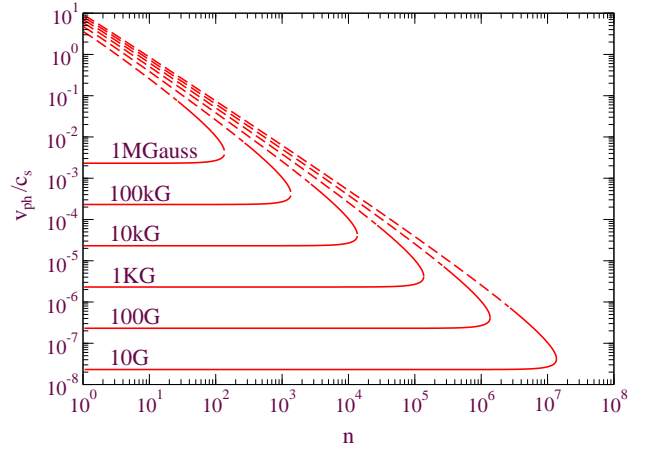} 
\caption{Plot of the $v_{ph}/c_s = \omega_1/k_x c_s$
against mode number, $n$, where $c_s$ is the adiabatic sound speed
and $\omega_1$ is the real part of the eigenfrequencies, $\omega =
\omega_1 (1 + i d)$, as predicted by Eq.~\pref{chieq}. Different
curves correspond to different background magnetic field
strengths, ${B}_0$. Solid lines represent resonances which are
inside the radiative zone, whilst dashed lines correspond to
unphysical modes whose resonances lie outside of the radiative
zone.\label{spectrumfig1}}
\end{figure}

\begin{figure}
\includegraphics[width=\columnwidth]{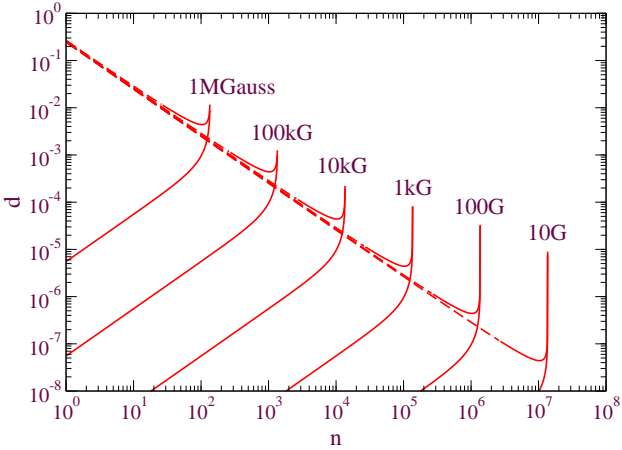} 
\caption{Plot of the $d$ against mode number,
$n$, where $d$ gives the imaginary part of the eigenfrequencies,
$\omega = \omega_1 (1 + i d)$, as predicted by Eq.~\pref{chieq}.
Different curves correspond to different background magnetic field
strengths, ${B}_0$. Solid lines represent resonances which are
inside the radiative zone, whilst dashed lines correspond to
unphysical modes whose resonances lie outside of the radiative
zone.\label{spectrumfig2}}
\end{figure}

Figures~\ref{spectrumfig1} and~\ref{spectrumfig2} present our
numerical solution of the eigenvalue spectrum obtained from
Eq.~(\ref{chieq}). Figure~\ref{spectrumfig1} plots $v_{ph}(n)/c_s
= \omega_1/(c_s k_x)$ {\it vs} mode number $n$ for various
magnetic fields, $B_0$. Figure \ref{spectrumfig2} similarly plots
$d$ against mode number for the same magnetic fields. In both
plots the parameter $\alpha=K_x/K_{\perp}$ is taken to be unity,
$\alpha=1$. The result with $\alpha \ne 1$ is trivially found by
rescaling $B_0$ because Eq.~(\ref{chieq}) depends only on the
product $\alpha B_0$. In both figures a dashed line is used if the
resonance of interest occurs above the top of the radiative zone
(and so outside the domain of many of our approximations).

Both of the figures~\ref{spectrumfig1} and~\ref{spectrumfig2} show two
branches of solutions up to a maximum mode number, as expected. Their
dependence on $n$ can also be understood analytically if we take $\mid
\chi\mid\gg \xi_c\sim 1$, and so $\eta \ll 1$. In this limit we obtain
from~\pref{mainspectrum1} the following approximate {\sl equation for
the MHD spectrum:}
\begin{eqnarray} \label{superloweq}
    &&\frac{\omega_1}{N}=\frac{2K_{\perp}}{\pi n} \,
    \left(z_r/2H + \ln 2\right)~,\nonumber\\
    &&d = \frac{\ln|\tan (\pi/\gamma)|}{2\pi n\left(1 - (z_r/2H + \ln
    2)^{-1}\right)}~,
\end{eqnarray}
which leads to the asymptotic solution
\begin{eqnarray}
\label{superlow}
&&\frac{\omega _{1}}{N}=\frac{2K_\perp}{\pi n}
\ln\left(\frac{4NH}{\alpha v_{Ac}}\frac{1}{\pi n}\right),\nonumber\\
&&d=\frac{\ln \left|\tan (\pi/\gamma )\right|}{2\pi n}~.
\end{eqnarray}
These modes correspond to the branches of the
figures~\ref{spectrumfig1} and~\ref{spectrumfig2} for which both
$v_{ph}/c_s$ and $d$ fall with $n$. The other branch can be found by
using the approximation $|\chi| \ll 1$ and hence $\cosh\chi \approx
\allowbreak 1+\chi^{2}/2$. The spectrum in this case is
\begin{eqnarray}
    &&\omega _{1}=\frac{\alpha K_{\bot } v_{Ac}}{H},\nonumber\\
    &&d=-\left( \frac{\alpha v_{Ac}}{4NH}\right) ^{2}  \pi n \ln \left| \tan
    \frac{\pi }{\gamma }\right|~.
\end{eqnarray}
Note that for this branch $v_{ph}(n)/c_s$ is independent of the
mode number, $n$, and $d$ grows with $n$, as is also seen in the
figure.

In addition to requiring the resonance to occur in the radiative
zone (the solid line in the figures), the validity of our
approximations also demand the frequency to not be smaller than
$10^{-5}$~s$^{-1}$ due to our neglect of the 27-day period solar rotation.
Inspection of Fig.~\ref{spectrumfig1} shows that these two
conditions are consistent with one another for a reasonably large
range of modes only for magnetic fields larger than a kG or so.

\section{The Resonant Properties}
\vskip 0.3cm
Several striking features emerge from the previous section's
analysis of the hydrodynamic eigenvalues and eigenfunctions. The
main two of these are:
\begin{enumerate}
\item The eigenfrequencies predicted by Eq.~\pref{mainspectrum1} are
complex, implying both damped and exponentially-growing modes, and
\item
The eigenmodes are not smooth as functions of $z$ about the
singular resonant point, $z=z_r(n,k_x,k_y)$, whose position
depends on the quantum numbers of the mode in question.
\end{enumerate}
These features lie at the root of the surprisingly large effects
which are produced by even moderate magnetic fields. We have
argued that both of these features can be traced to a resonance
between $g$-modes and Alfv\'en waves, which arises due to the
existence of gradients in the background density, $\rho_0(z)$.

In this section we expand on these two properties, to more
precisely pin down their origin. In particular, we expect the
development of complex frequencies to indicate a hydrodynamical
instability. This is much like the imaginary part which $g$-waves
develop as one passes into the convective zone from the radiative
zone. For $g$-modes this imaginary part arises because buoyancy no
longer acts as a restoring force in the convective zone, but
instead becomes destabilizing. Indeed the instability in this case
drives the convection, which is why the radiative zone ends. We
expect a similar instability to arise for the Alfv\'en/$g$-mode
resonance.

\subsection{Instability}
The necessity for complex frequencies is the most disturbing
feature about the above eigenvalue problem, because it implies
mode functions which grow exponentially in amplitude with time.
This signals an instability in the physics which pumps energy into
these resonances, and this section aims to discuss the nature of
this instability, and the physical interpretation to be assigned
to the imaginary part, $d$.

Exponentially-growing instabilities within an approximate ({\it
e.g.} linearized) analysis reflect a system's propensity to leave
the small-field regime, on which the validity of the approximate
analysis relies. The question becomes: where does the instability
lead, and what previously-neglected effects stabilize the runaway
behaviour?

In the present instance we shall see that the normal modes become
strongly peaked near the resonant radii, and that the energy flow
near these radii is directed along the resonance plane, as would
be true for a pure Alfv\'en wave. This leads us to the following
picture. Since helioseismic waves are likely generated by the
turbulence at the bottom of the convective zone, it is natural to
imagine starting the system with a regular helioseismic $g$-mode
and asking how it evolves. In this case the resonance allows the
energy in this mode to be funnelled into the Alfv\'en wave, and so
to be channelled along the magnetic field lines away from the
solar equatorial plane. This process drives us beyond the range of
our approximations by making the wave amplitude large near the
resonant surface, but also by driving energy out of the equatorial
plane where the rectangular analysis applies.

We arrive in this way at an interpretation in which the
instability signals the excitation of Alfv\'en waves from
$g$-modes due to their resonant mixing. In this picture the
imaginary part of the frequency, $\Gamma = \Im \, \omega = d \,
\omega_1$, describes the rate at which the Alfv\'en mode is
excited in this process. Since the mixing and excitation are
weak-field hydrodynamic phenomena it is reasonable that this rate
can be computed using only the linearized magneto-hydrodynamical
equations.

Once excited, the mode amplitudes near the resonances grow until
the energy in them becomes dissipated by effects which are not
captured by the approximate discussion we present here. The rate
for this dissipation must grow as the mode amplitude grows, until
it equals the production rate, $\Gamma$, at which point a steady
state develops and the mode growth is stabilized. Because the
resonant mode grows until its damping rate equals its production
rate, it suffices to know the mode's growth rate in order to
determine the overall on-resonance amplitude.

How much more can be said about the amplitude of the modes depends
on whether or not the mode stabilizes at small enough amplitudes
to allow the hydrodynamic approximation to the production rate,
$\Gamma = d \, \omega_1$, to be accurate. For instance, if the
mode only stabilizes once it is large enough to require a
nonlinear analysis, then the final production and dissipation
rates may be very different from the linearized expressions
derived above. If, on the other hand, the modes saturate at
comparatively small amplitudes by dissipating energy into
non-hydrodynamical modes (such as by Landau damping), then it can
happen that the stabilized mode amplitude is not large enough to
significantly change the linearized prediction for its production
rate.

In what follows we use the imaginary part of $\omega$ to determine
the resonant mode's amplitude at the position of the resonance.
There is no loss in doing so to the extent that we take $\Im
\omega$ as a parameter which is not related to the mode quantum
numbers by the hydrodynamical equations of motion. On occasion we
shall also use the explicit formulae for $\Im \omega$ which is
predicted by the hydrodynamics, and in this case we are assuming
that the mode stabilization occurs at small enough amplitudes to
not invalidate the assumption that this is a good approximation to
the mode production rate.

\subsection{Spatial dependence}
We next address the singular shape of the wave fronts at the
resonant points, $z = z_r$. Mathematically, this singularity
arises because the interval of physical interest, $0 \le z \le
z_*$, corresponds to the interval $0< \xi_c \le \xi \le \xi_*$,
and this contains the singular point $\xi =1$ provided $\xi_c < 1$
and $\xi_* > 1$. The hypergeometric functions can -- and do --
develop singularities at $\xi = 1$, and it is this singular
behaviour which we now examine.

Using Eq.~(\ref{xilarge1}) of the Appendix~B one finds as $\xi \to
1$ (from either side):
\begin{equation}
\label{singularity}
    b_z \propto |\xi - 1|^{1-\nu} + \hbox{nonsingular}
    \quad \hbox{and} \quad \frac{db_z}{dz} \propto
    |\xi - 1|^{-\nu},
\end{equation}
with $\nu = (\gamma - 1)/\gamma$ (giving $\nu = \frac25$ for an
ideal gas, for which $\gamma = \frac53$). For $0 < \nu < 1$ we see
that $b_z$ remains finite but $db_z/dz \to \infty$ as $\xi \to 1$.
As we shall see, once the eigenfrequency's imaginary part is
included $db_z/dz$ does not actually diverge as $\xi \to 1$, but
instead adopts a systematically large resonant line-shape.

Using this information in eqs.~\pref{all} allows the asymptotic
form for all other quantities to be determined at the resonant
point. In particular, since $v_z$ does not involve $db_z/dz$, it
also remains finite on resonance, unlike all other components of
the fluid velocity. This indicates that it is the transverse
velocity perturbations (and so also the transverse kinetic energy
of the oscillations) which increase most dramatically at the
resonant plane.

\ssubsubsection{Positions of resonances}
The equation for the position of resonances can be found by using the
resonance condition
\begin{equation}
\label{rexi} \hbox{Re}~\xi = \hbox{Re}~\xi_c \; e^{z/H} =
\left(\frac{k_xv_{Ac}}{\omega_1(n)}\right)^2 e^{z_r(n)/H} = 1
\end{equation}
and Eq.~\pref{superloweq}. It takes the following transcendental form
\begin{equation}
\label{positioneq}
z_{s}(n) = H \left[ \frac{K_{x}}{K_{\perp }} \left( \frac{\pi | n|
\, v_{Ac}}{c_{s}} \right) \frac{\gamma }{\sqrt{\gamma -1}}\;
e^{z_{s}(n)/2H}-\ln 4\right]\,.  
\end{equation}
Taking into account the asymptotic form for the real part of the
eigenfrequency, $\omega_1(n)$, as given by Eq.~\pref{superlow}, leads
to the following asymptotic expression for the dependence of the
resonant layer position on the node number, $n$:
\begin{equation}
\label{position}
z_r(n)=2H\ln\left[\frac{4NH}{\alpha v_{Ac}}\frac{1}{2\pi n}
\ln\left(\frac{4NH}{\alpha v_{Ac}}\frac{1}{\pi n}\right)\right]\,,
\end{equation}
here
\begin{equation}
v_{Ac}= \frac{B_0}{\sqrt{4\pi \rho_{c}}} = \left(
\frac{B_0}{43.4~\hbox{G}}\right) \; \hbox{cm s}^{-1}
\end{equation}
is the Alfv\'en velocity at the solar center.

\begin{figure}
\includegraphics[width=\columnwidth]{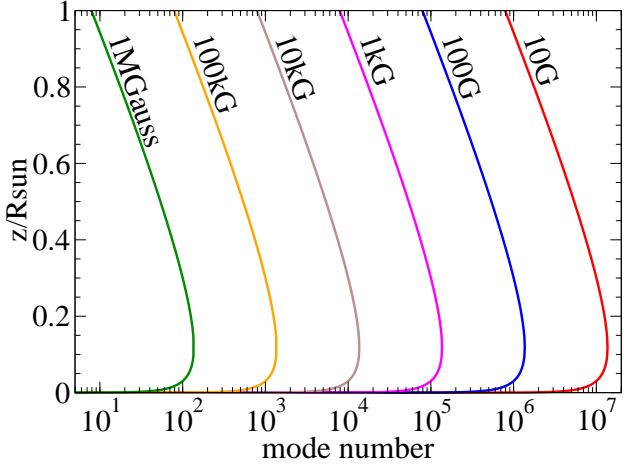} 
\caption{A plot of the resonant position,
$z_r(n)$, {\it vs} mode number $n$ for different magnetic fields
in the range $B_0 =$~10G--1MGauss.\label{resposfig}}
\end{figure}

Fig.~\ref{resposfig} plots the position, $z_r(n)$, predicted by
Eq.~(\ref{position}) for the case of longitudinal wave propagation
($k_y = 0$) and for magnetic fields in the range 
$B_0=$10G--1MGauss. (The same result for an obliquely-propagating wave with
$k_y \ne 0$ is produced by a larger value for $B_0$.) This plot is
the analog of Figs.~5a,b in~\cite{DzhalilovSemikoz}.

\ssubsubsection{Spacing of resonances}
Knowing the position of the resonances also permits us to
determine the distance between them. This quantity is relevant to
the propagation of particles like neutrinos through the resonant
waves. The spacing is:
\begin{equation}
    z_r(n +1) - z_r(n) = \left(\frac{K_x}{K_{\perp}}\right) \frac{\pi
    H \, v_{Ac}\gamma}{c_s\sqrt{\gamma -1}} \; e^{z_r(n)/2H} ~.
\label{distance}
\end{equation}

\begin{figure}
\includegraphics[width=\columnwidth]{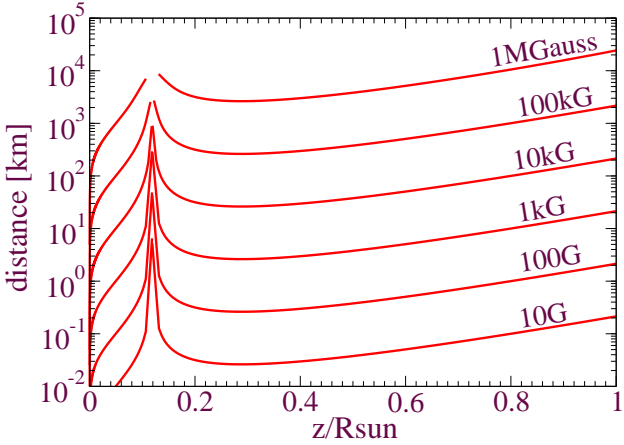} 
\caption{The distance between
neighbouring Alfv\'en resonant layers {\it vs} the position of the
layer within the solar interior. \label{resspacefig}}
\end{figure}

This dependence of this quantity on mode number, $n$, is shown in
Fig. \ref{resspacefig} for longitudinal wave propagation, $k_y=0$,
and for different magnetic field values. From this figure we see,
in the region $z_r\ga 0.3R_{\odot}$, that the distance between
layer positions grows with magnetic field and with distance from
the solar center. Both of these features were already seen in
preliminary WKB calculations of the resonances (see Figs. 7a, 7b
of ref.~\cite{DzhalilovSemikoz}), and our present numerics also
confirm that the spacing is approximately proportional to $B_0$
for $z \ga 0.3 R_\odot$, that was seen earlier.

Eq.~(\ref{distance}) can be simplified using the expression,
Eq.~(\ref{position}), for the resonance position:
\begin{equation}
\label{simple} z_r(n +1) - z_r(n) \approx \frac{2H}{\mid n\mid}~,
\end{equation}
where the inverse mode number $\mid n \mid^{-1}$ is proportional
to the background magnetic field $\sim B_0$. This is in accordance
with the spectrum Eq.~(\ref{superloweq}), for which the phase
velocity $v_{ph}=\omega_1/k_x$ equals the Alfv\'en speed at the
resonance position, $\omega_1/k_x = v_A(z_r,B_0) \sim B_0$.

Numerically, it is noteworthy that the spacing between resonances
can be in the ballpark of hundreds of kilometers. This is
significant because this is close to the resonant oscillation
length, $l_{osc}^{res}$, for $E\sim \rm{MeV}$ neutrinos, if
-- as now seems quite likely -- resonant LMA oscillations are
responsible for explaining the solar neutrino problem since
\begin{equation}
\label{nuosc} 
l_{osc}^{res}= \frac{250~\rm{km}~
(E/\rm{MeV})}{\Delta m^2_5\sin 2\theta}\,, 
\end{equation}
where $\Delta m^2_5 = \Delta m^2/10^{-5}~ \hbox{eV}^2$. Repeatedly
perturbing neutrinos over distance scales comparable to
$l_{osc}^{res}$ has long been known to be a prerequisite for
disturbing the standard MSW picture of oscillations in the solar
medium. This raises the possibility -- recently explored in more
detail in ref.~\cite{Burgess3} -- that $g$-mode/Alfv\'en
resonances can alter neutrino propagation. If so, the observation
of resonant oscillations of solar neutrinos may provide some
direct information about the properties of the MG waves we discuss
here.

\ssubsubsection{Widths of the resonance layers}
We next examine the energy flow in the vicinity of a resonance.
This gives some information about the nature of the instability
which is indicated by the complex mode frequencies. We find here
that there is a large energy flow along the magnetic field lines,
perpendicular to the $z$ direction. We use the dimension of the
region in which this energy flows as a measure of the width of the
resonant layer.

In the linearized regime the energy flux, $\vector{S}$, has both
mechanical and electromagnetic (Poynting) components, $\vector{S}
= \vector{S}_{\rm mech} + \vector{S}_{\rm em}$, with
\begin{eqnarray}
\vector{S}_{\rm mech} &=&  p' \, \vector{v}~,\\
\vector{S}_{\rm em} &=& \frac{1}{4 \pi} \,\left\{
\Bigl[\vector{B}'\times [\vector{v}\times \vector{B}_0] \Bigr]
\right\} .\nonumber
\end{eqnarray}
In order of magnitude, the relative size of these contributions is
$B_0 B'/(4 \pi p')$ and this turns out to be small. This smallness
is easy to see to the extent that $B'/B_0 \sim p'/p_0$, since then
$B_0 B'/(4 \pi p') \sim B_0^2/(4 \pi p_0)$, which is small by
virtue of the background inequality $B^2_0/(4\pi) = \gamma \,
(v_A^2/c_s^2) \, p_0 \ll p_0$.

Since the mechanical energy flux dominates, its direction is
simply given by the direction of $\vector{v}$ near the resonance.
This may be determined using the asymptotic form as $\xi \to 1$ in
the expressions for $p'$ and $v_z$ in eqs.~\pref{all}. Notice that
$v_z$ approaches a finite value in this limit, while $v_x, v_y$
and $p'$ all diverge as $\xi \to 1$. This shows that the energy
flow, like the fluid flow, is mainly parallel to the resonant
plane when evaluated near a resonance.

Using the asymptotic expressions, $db_z(z)/dz \sim {\cal B} \;
|1-\xi|^{-\nu}$, and  $b_z(z) \sim {\cal B} \;|1-\xi|^{1-\nu} +
{\cal C}$, with constants ${\cal B}$, ${\cal C}$ and $\nu =
(\gamma - 1)/\gamma \approx 0.4$, we obtain $v_z \propto |b_z|
\sim {\cal B} \, |1 - \xi|^{1-\nu}$, and
\begin{eqnarray}\label{pvxvy}
&&p' , v_x, v_y \propto \left|\frac{db_z}{dz}\right| \sim
\frac{|{\cal B}|}{|1-\xi|^\nu} =\\
&&= \frac{|{\cal B}|}{|1-\xi_r+\xi_r2id|^\nu}~ = 
\frac{|{\cal B}|}{[(1-\xi_r)^2 +
4\, \xi_r^2 d^2|^{\nu/2}}~,
\nonumber
\end{eqnarray}
where we have used our definitions -- $\omega =\omega_1(1+id)$,
with $|d|\ll 1$, and so $\xi = \xi_{r}(1 - 2id)$, with $\xi_{r} =
\Bigl({k_xv_{Ac}/\omega_1} \Bigr)^2e^{z_r/H}$ -- to write
the real and imaginary parts explicitly. The energy flow near the
resonant plane is similarly given by the time average $\vector{S}
= \langle p' \vector{v} \rangle= \frac12 {\rm
Re}(p'\vector{v}^*)$, where the symbol $\langle \cdots \rangle$
means average over time. Combining the results for $p'$ and
$\vector{v}$ we see that near resonance
\begin{eqnarray}
 S_z &\propto& \frac{1}{[(1-\xi_r)^2 + 4\, \xi_r^2 d^2]^{\nu/2}}~,\nonumber\\
 S_x, S_y &\propto& \frac{1}{[(1-\xi_r)^2 + 4\, \xi_r^2
d^2]^{\nu}}~
\end{eqnarray}
and so, for instance, $S_z$ reaches half of its resonant value
when $\xi = \xi_\pm = 1 \pm \delta \xi$ where $[(1-\xi_\pm)^2 +
4\, \xi_\pm^2 d^2]^{\nu/2} = 2 (2d)^{\nu}$.

On the other hand, for $z$ well beyond the resonant zone, where
$\xi > 1$, we know that $b_z(z)$ is given by
Eq.~(\ref{evanescent}), which states $b_z(z) \propto
\xi^{-K_{\perp}} F(a,a^*;1+2K_{\perp};\xi^{-1})$. By virtue of the
condition $a^*=1 -c +a$ we see that this is a real function (in
the limit $d\to 0 $). Consequently, the explicit factors of `$i$'
in eqs.~(\ref{all}) imply that quantities like the pressure
perturbation, $p'(z)$, and the transverse velocities $v_{x,y}(z)$
are pure imaginary. It therefore follows that the $z$-component of
the energy flux vanishes in the region $z>z_r$, outside of the
resonance, because $S_z= \frac12 {\rm Re}(p' v_z^*)=0$, while the
transverse components are nonzero in this region, but are falling
off exponentially.

This peaking, then vanishing, of $S_z$ motivates our choice of the
width of the $S_z(z)$ resonant line-shape as a measure of the
resonance layer width. {}From this condition we can easy find the
results for the widths $z_\pm$ (or, equivalently, $\xi_\pm$):
\begin{equation}
\delta \xi^2 = (2|d|)^2 \, \left(2^{2/\nu} - 1 \right)~,
\end{equation}
giving our final expression for the width of the Alfv\'en singular
layer:
\begin{equation}
\delta z =z_+ - z_- = H \ln\left( {\frac{\xi_+}{\xi_-}} \right)
\approx 4H|d|2^{\gamma/(\gamma-1)}~. \label{width}
\end{equation}
Substituting the expression for $d$ in terms of mode number, using
Eq.~(\ref{superlow}), this becomes
\begin{equation}
\Delta z_r(n)\approx \frac{2H}{\pi n}2^{\gamma/(\gamma
-1)}\ln\left| \tan \frac{\pi}{\gamma} \right| ~.
\end{equation}

\begin{figure}
\includegraphics[width=\columnwidth,height=0.5\columnwidth]{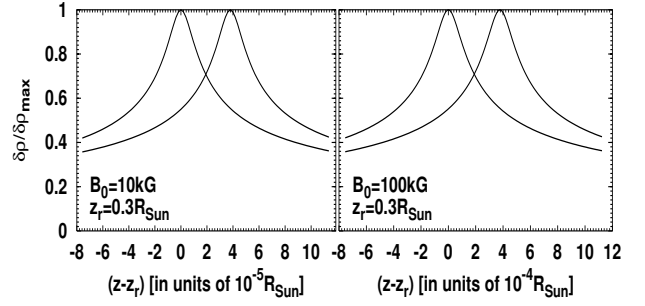} 
\caption{A plot of neighboring density
profiles, showing their width in comparison with the spacing in
between them. The plot is done for resonances in the region
$z_r\sim 0.3R_{\odot}$.\label{overlapfig}}
\end{figure}

Together with Eq.~(\ref{simple}) we then get the ratio of the
distance between resonances to the width of each resonance,
\begin{equation}
\label{noisestructure} \frac{z_r(n +1) - z_r(n)}{\Delta
z_r(n)}=\frac{\pi}{2^{\gamma/(\gamma -1)}\ln\mid \tan
\frac{\pi}{\gamma} \mid}\approx 0.5~.
\end{equation}
Figure~\ref{overlapfig} plots the density profiles of two neighboring
resonances in the region $z_r\sim 0.3R_{\odot}$. In this region their
width is approximately twice their separation as measured by the
distance between peaks. That is not surprising since the shape of the
density profile is the same as in Eq.~\pref{pvxvy} for other MHD
perturbations (see below, Eq.~\pref{deltarhospike}).

\ssubsubsection{Density Profiles}
We now give the explicit expression for the density profile predicted
for our wave solutions. The basic expression is obtained through the
direct substitution of our expression for $b_z(z)$ -- keeping in mind
eqs. (\ref{solution}), (\ref{ratio1}) and (\ref{notation}) -- into
the last of eqs.~\pref{all}. We find in this way the following
expression for $\rho'/\rho_0$, divided by the arbitrary wave
normalization, $D_1$:
\begin{eqnarray} 
\label{MHDdensity}
&&\frac{1}{D_1}\left( \frac{\rho' (z)}{\rho_0(z)} \right)=\nonumber\\
&&= \frac{-i\xi_c^{\sigma_1}}{k_xH} \left\{ \left( \frac{\gamma -
1}{\gamma} + \frac{\sigma_1\omega^2}{k_{\perp}^2c_s^2}\right)
\left(
\frac{\xi}{\xi_c}\right)^{\sigma_1}F(a_1,b_1;c_1;\xi ) \right. \nonumber\\
&&- \left( \frac{\gamma - 1}{\gamma} +
\frac{\sigma_2\omega^2}{k_{\perp}^2c_s^2}\right)\Bigl
(\frac{\xi}{\xi_c} \Bigr )^{\sigma_2}F(a_2,b_2;c_2;\xi )
\frac{F(a_1,b_1;c_1;\xi_c ) }
{F(a_2,b_2;c_2;\xi_c )} \nonumber\\
&& + \frac{\xi_c\omega^2}{k_{\perp}^2c_s^2}
\left[\frac{(\sigma_1^2 - K_{\perp}^2)}{1 + i\beta} \left(
\frac{\xi}{\xi_c} \right)^{\sigma_1 +
1}F(a_1 +1,b_1 +1;c_1 +1;\xi) \right. \nonumber\\
&&  - \frac{(\sigma_2^2 - K_{\perp}^2)}{1 - i\beta}
\; \left(\frac{\xi}{\xi_c} \right)^{\sigma_2 + 1} F(a_2 +1,b_2
+1;c_2 +1;\xi) \times\nonumber\\
&&\left.\left.\times \frac{F(a_1,b_1;c_1;\xi_c ) }
{F(a_2,b_2;c_2;\xi_c )} \right] \right\}\,, 
\end{eqnarray} 
where $\sigma_1$ is given by Eq.~(\ref{sigmadef})
for $q=-i\beta$, $\sigma_2=(\gamma - 1)/2\gamma - i\beta/2$;
$a_{1,2}=\sigma_{1,2} + K_{\perp}$, $b_{1,2}=
\sigma_{1,2}-K_{\perp}$; $c_{1,2}= 1 \pm i\beta$ (see Eq.~(\ref{notation})).

The behaviour of the density perturbation near the resonant point,
$z_r$, takes a simpler form for the Lagrange perturbation
\begin{eqnarray}
&&    \delta \rho(\vec{r}) = \rho'(\vec{r}) + \delta \vec{r}\cdot
    \nabla \rho_0 = \nonumber\\
&&=[\rho'(z) - (\delta z/H)\rho_0(z)]e^{-i(\omega t
    -k_xx -k_yy)},
\label{deltarhospike}
\end{eqnarray}
where the Eulerian perturbation $\rho'(z)$ is given by
Eq.~(\ref{MHDdensity}). In linear MHD the Lagrangian density
perturbation is directly constrained by mass conservation,
Eq.~(\ref{mass}), $-i\omega \rho_0^{-1}\delta \rho (z)= -u$, where
$u = \div \vector{v}$ is the compressibility, given by
Eq.~(\ref{u1}). One obtains in this way the following Lagrange
density perturbation near the resonance, $z\sim z_r(n)$:
\begin{equation} 
\label{Lagrange}
    \frac{1}{D_1}\frac{\delta \rho (z)}{\rho_0} \approx C
     -\left[e^{-(z - z_r(n))/H}(1 + 2id) -1\right]^{-{(
     \gamma -1)}/{\gamma}}\,,
\end{equation}
where 
\[
C =\frac{\Gamma (\gamma^{-1})}{\Gamma(-\gamma^{-1})}
\left(
\frac{\gamma v_{ph}(n)}{i\sqrt{\gamma -1} \; c_s}
\right)^{2/\gamma}
\left(\frac{\gamma -1}{4\gamma} + \frac{\gamma^2K_{\perp}^2}{\gamma -1} \right)\,.
\]
Here the phase velocity, $v_{ph}(n)=\omega_1(n)/k_x$, is given by
the solution to the eigen-spectrum, $\omega=\omega_1(1 + id)$,
such as obtained in Eq.~(\pref{superlow}), and we include the
unknown normalization factor $D_1$.

This quantity $\left| \delta \rho/\rho_0\right|$ is plotted in
Fig.~\ref{overlapfig} for two neighbouring resonances near $z =
0.3R_{\odot}$ for the two values of the magnetic field, $B_0=$10kG and
$B_0=$100kG. We see that the presence of the resonances introduces a
series of density excursions, located at the resonant positions. The
disappearance of the resonant density spikes in the zero-field limit
is most easily seen by noticing that in the limit $B_0\to 0$ the
position $z_r$ becomes negative, $z_r=-H\ln 4<0$ {\it (see
Eq.~\pref{positioneq})}, or occurs outside the physical region $0\leq
z\leq z_*$.

\section{Possibilities for Observations}
Given the strength of the resonant effects on the helioseismic
wave profiles, we next explore some of their observable properties
to determine how they might be detected, and to what extent their
effects are consistent with current helioseismic observations.

\subsection{{Measurements of Mode Frequencies?}}
Many helioseismic waves have been detected, but to date all of these
are pressure-driven $p$-modes rather than the usually-lower-frequency
$g$-modes. The latter are more difficult to detect because they are so
strongly attenuated within the convective zone.

Because the central Alfv\'en frequency is so small compared with
the other helioseismic frequencies, we only expect resonances
between Alfv\'en waves and those helioseismic waves which
penetrate most deeply into the solar interior, and which have the
lowest frequencies. This is why the resonance we have found
involves mixing only with $g$-modes, and not $p$-modes. It follows
that so long as observations are restricted to $p$-modes, the
resonances we describe here will have negligible implications for
the comparison between measured $p$-wave frequencies and solar
models.

Things become more interesting should $g$-modes be observed, since
then a comparison between observations and predicted frequencies
could depend more strongly on the solar magnetic field. Figure
\pref{spectrumfig1} (or Eqs.~\pref{superlow} and \pref{position})
show how the frequencies of the resonant modes depend on magnetic
field. Unfortunately, frequencies this low are likely to be a
challenge to detect for quite some time.

Of course, a proper comparison requires the inclusion of the full
(spherical) geometry of the problem, as well as temperature and
gravity gradients, solar rotation, and other factors which we ignore
here. Our goal in this paper is merely to point out the potential for
larger-than-expected effects, and to obtain a rough first
determination of the sensitivity to magnetic fields.

\subsection{Constraints on Mode Amplitudes}
Mode amplitude remains undetermined within the linear approximations
assumed here, and is ultimately controlled by the mechanisms which
stimulate and damp the waves of interest in the solar environment.
Amplitude can nevertheless be characterized in any of three equivalent
ways: the size of the mode's contribution to the mass-density profile,
the size of the fluid velocity, $v_z$, at the solar surface, and the
total mode energy.  In principle, our solutions can be used to express
any two of these in terms of the third, as functions of the background
magnetic field.

A sufficiently detailed comparison of the resulting expressions --
should this become possible -- would then allow one to obtain the
magnetic field strength. We illustrate this by calculating here
  the energy content of the resonant MG waves.

\ssubsubsection{Mode Energy}
The mode energy is a useful way to quantify the mode amplitude. Here
we provide formulae which make this connection explicit.  The mode
energy is most simply computed by using the virial theorem, which
relates the total energy to the kinetic energy averaged over a single
period of oscillations: $E = 2 \langle E_{\rm kin} \rangle$:
\begin{equation}
    \langle E_{\rm kin} \rangle = \frac12 \int d^3\vector{x} \; \rho_0 \,
    \langle [\hbox{Re}~\vector{v}(\vector{x},t)]^2\rangle ~.
    \label{kinenergy}
\end{equation}
For modes having a small imaginary frequency, $| d| \ll 1$, one
obtains the following simple approximate formula for the above
average
\begin{eqnarray}
\langle [\hbox{Re}~\vector{v} (\vector{x},t) ]^2 \rangle &=& \frac{1}{2}
    \left| \vector{v}(z) \right|^2 \\ 
&& +\frac{1}{4} [ (\hbox{Im}~v_i(z))^2 -
    (\hbox{Re}~v_i(z))^2] \nonumber\\
&&\times\sin [2(k_x x + k_y y)] + \hbox{O}(d)~,\nonumber
\end{eqnarray}
The second term in this expression is typically negligible once it
is integrated over $dx$ and $dy$.

Using our low frequency approximation, $\omega^2\ll k_x^2c_s^2$,
we find the following result for the squared velocity,
$|\vector{v}(z)|^2= |v_z|^2 + | v_x|^2 + | v_y|^2$:
\begin{eqnarray}\label{velocity}
&&|\vector{v}(\xi)|^2 = v_{ph}^2 \left\{ |b_z(\xi)|^2
    + \frac{\alpha^2}{K_x^2} \left( \frac{|b_z(\xi)|^2}{\gamma^2} -\right.\right.\\
&&\left.\left.\frac{1}{\gamma} \left[ b_z^*(\xi) \xi \frac{db_z(\xi)}{d\xi} +
    b_z(\xi) \left( \xi \frac{ db_z(\xi)}{d\xi} \right)^* \right]+ \left| \xi \frac{db_z(\xi)}{d\xi} \right|^2 \right)
    \right\}\nonumber 
\end{eqnarray}
We next integrate this result over the solar volume. In our simplified
geometry we do so by bounding the transverse directions with a circle
of radius $R_\odot$, having cross-sectional area $\pi R_{\odot}^2$.
This gives the basic expression for the energy per unit length as
$dE/dz = (\pi/4) R_{\odot}^2 \rho_0(z) | \vector{v}(z) |^2$.

A useful expression is obtained by eliminating the unknown
normalization, $D_1$ between eqs.~\pref{velocity} and
\pref{MHDdensity}. This gives a direct, but complicated, relation
between the mode energy and the size of the peak density excursion on
resonance. To this end we denote the function in brackets in
Eq.~\pref{velocity} by $A(\xi)$. With this choice we have
$|\vector{v}|^2 = v_{ph}^2(n)A(\xi,n)$, with $v_{ph}^2(n)=
\omega^2/k_x^2= v_A^2(z_r)= B_0^2 e^{z_r(n)/H}/(4\pi \rho_{c})$.  Note
that this last expression uses the resonance condition, $\xi =
(k_xv_A(z_r))^2/\omega^2 =1$, for the mode eigenfrequencies. If we
similarly denote the right-hand-side of Eq.~\pref{MHDdensity} by
$B(\xi)$, so $\rho'/\rho_0 = D_1 \, B(\xi)$, then we find
\begin{eqnarray}
&&E_{MHD}(n) = E_0\left(\frac{B_0}{1~\hbox{Gauss}}\right)^2\times
\frac{1}{(\hbox{Re}~B(\xi=1,n))^2 }\times\nonumber\\
&&\times\left(\frac{\hbox{Re}~
\rho'(z_r(n))}{\rho_0(z_r(n))}\right)^2\times
\int_{\xi_c}^{\xi_{*}}\frac{d\xi }{\xi^2}
\frac{A(\xi,n)}{D_1^2(n)} ~ , \label{energy}
\end{eqnarray}
where for convenience we assume $D_1$ to be real. In this
expression we have used the explicit background density
distribution, $\rho_0 = \rho_{c} \, e^{-z/H}$, and changed
variables from $z$ to $\xi$. The lower limit of integration is
$\xi_c = e^{-z_r(n)/H}$ which takes values $1 > \xi_c \ga
10^{-3}$ for a resonance between the solar center and the bottom
of the convective zone. We take as upper limit $\xi_{*} = e^{(0.7
\, R_{\odot} - z_r)/H} \gg 1$ for the same choice of the Alfv\'en
resonance positions.

The energy scale, $E_0$, in Eq.~\pref{energy} denotes the combination
$E_0 = (1/16) R_{\odot}^2H \times (1~ \hbox{Gauss})^2$. Using
$(1~\hbox{Gauss})^2 = 1$ erg/cm${}^{-3}$ gives $E_0 \sim 2.1\times
10^{30}$ erg. From this we can see that, for magnetic fields
$B_0=10^5-10^4$ Gauss, density spikes as large as $\rho'/\rho_0 \sim
1-10$ per cent will involve energies $E \lsim 10^{36}$ erg, provided
$A/D_1^2$ and $B$ are order unity, which is reasonable for the maximum
g-mode energy.

\ssubsubsection{Density Perturbations and Neutrino Oscillations}
The g-mode/Alfv\'en resonances could give rise density perturbations
with a large amplitude. Are such density excursions constrained by
observations? At first sight one might think so, because the success
of the comparison of standard solar models with helioseismic
observations constrains the solar density profile to be within 1~per
cent of solar model predictions. Since these models ignore magnetic
fields, one might expect magnetic-field-induced effects like those we
consider here to be ruled out at this level.

Unfortunately, deconvoluting solar properties from observed wave
profiles is at present only possible subject to smoothness
assumptions. As a result current constraints on deviations of $\rho$
from solar model predictions only apply to profiles which do not vary
on scales shorter than around 1000 km
\cite{Castellani:1997pk,Christensen,Christensen-Dalsgaard:2002ur}.
It is for this reason that present analyses do not yet exclude the
existence of such resonances at an interesting level.
Observable effects are only possible if the distance between spikes is
comparable to the neutrino oscillation length at the position where
neutrino conversions occur.  Furthermore, their amplitude must be of
order a few percent in order to produce observable effects.

If these density spikes are sufficiently large, they could have
implications for neutrino oscillations, and we have argued
\cite{Burgess3} that this may be the first place where they are
potentially detectable.  Besides identifying whether the required
density profiles can be obtained using reasonable magnetic fields, a
proper analysis must also ask whether the resulting modification of
the signal can be identified within the considerable uncertainties
which are inherent in any neutrino measurement.
 
The sensitivity of solar neutrino oscillations to noise in the solar
interior has been re-examined in ref.~\cite{Burgess3}, using the best
current estimates of neutrino properties. The outcome is that the
measurement of neutrino properties at KamLAND now provides potentially
important information on fluctuations in the solar environment on
scales $l_{osc}^{res}$ (see Eq.~\pref{nuosc}) to which standard
helioseismic constraints are currently insensitive.

Performing a combined fit of KamLAND and solar data it has also been
shown how the determination of neutrino oscillation parameters
strongly depends on the magnitude of solar density fluctuations. 

The Alfv\'en/$g$-mode resonances turn out to have the right spacing
and position for magnetic fields in the range of $B_0 =10-100$~kG.
Depending on the size of the magnetic fields which are ultimately
found in the radiative zone, this perhaps opens a completely new and
interesting observational window on the solar interior.

\section{Discussion}
Here we have presented an analytic discussion of the influence of
magnetic fields on helioseismic waves in a simplified geometry. We
have argued that this geometry captures the physics not too close to
the solar center ($r \ga H \sim 0.1 R_\odot$) within the radiative
zone, provided that the background magnetic field is perpendicular to
the background density gradient. In particular it would apply to a
dipole magnetic field configuration near the dipole's equatorial
plane.

We find that within our approximations sufficiently large magnetic
fields can cause significant changes to the profiles of helioseismic
$g$-waves, while not appreciably perturbing helioseismic $p$-waves.
The comparatively large $g$-wave effects arise because of a resonance
which occurs between the $g$-modes and magnetic Alfv\'en waves in the
solar radiative zone. Although the radiative-zone magnetic fields
required to produce observable effects are larger than are often
considered -- more than a few kG -- they are not directly ruled out
by any observations.  Noting and characterizing the basic features of
such resonance constitutes our main result.

Although the density profiles at their maxima could have amplitudes as
large as a few percent or more on resonance, we do not believe the
corresponding radiative-zone magnetic fields can yet be ruled out by
comparison with helioseismic data, since the density excursions are
sufficiently narrow (hundreds of kilometers) as to evade the
assumptions which underlie standard helioseismic analyses.

For magnetic fields in the 10 kG range, the best hopes for detection
of the resonant waves may be through their influence on neutrino
propagation. This influence essentially arises because the presence of
strong density variations affects the solar neutrino survival
probability, with a corresponding change in the resulting solar
neutrino fluxes.  As shown in ref.~\cite{Burgess3}, the measurement
of neutrino properties at KamLAND provides new information about
fluctuations in the solar environment on correlation length scales
close to 100~km, to which standard helioseismic constraints are
largely insensitive. It has also been shown how the determination of
neutrino oscillation parameters from a combined fit of KamLAND and
solar data depends strongly on the magnitude of solar density
fluctuations. 

Since the resonances rely on the condition that $\vector{B} \perp
\nabla \rho$, there are several magnetic-field geometries to which
our analysis might apply, of which we consider here two
illustrative extremes. Suppose first that, in spherical
coordinates $(r,\theta,\phi)$, we imagine $B_r \approx 0$ but
$B_\theta \ne 0$. Then the field is always perpendicular to a
radial density gradient and the resonance we find might be
expected to arise in all directions as one comes away from the
solar center. In this case the solar $g$-modes would tend to be
trapped behind the resonance, and so are kept away from the solar
surface even more strongly than is normally expected. This would
make the prospects for their eventual detection very poor.

Alternatively, if the magnetic field has more of a dipole form it
might be imagined that the condition $\vector{B} \perp \nabla \rho$
only holds near the solar equatorial plane, and not near the solar
poles. In this case a more detailed calculation is needed in order to
determine the resulting wave form. This kind of geometry could have
interesting consequences for the solar neutrino signal, because in
this case the deviation from standard MSW analyses only arises for
neutrinos which travel sufficiently close to the solar equatorial
plane. Given the roughly 7-degree inclination of the Earth's orbit
relative to the plane of the solar equator, there is a possibility of
observing a seasonal dependence in the observed solar neutrino flux.
Since the presence of the MG resonance tends to decrease the MSW
effect, the prediction would be that the observed rate of solar
electron-neutrino events is maximized when the Earth is closest to the
solar equatorial plane (December and June) and is minimized when
furthest from this plane (March and September).

\section*{Acknowledgements}

This work was supported by Spanish grant BFM2002-00345, by European
RTN network HPRN-CT-2000-00148, by European Science Foundation network
grant N.~86, by Iberdrola Foundation (VBS), by INTAS YSF grant
2001/2-148 and MECD grant SB2000-0464 (TIR).  C.B.'s research is
supported by grants from NSERC (Canada) and FCAR (Quebec).  VBS, NSD
and TIR were partially supported by the RFBR grant 00-02-16271.  C.B.
would like to thank hospitality in Valencia during part of this work.

\section*{Appendix A. Some Hypergeometric Facts}

We here record some properties of Hypergeometric functions which
are useful in obtaining our results. The series solution to the
Hypergeometric equation, Eq.~\pref{Gauss}, which is analytic throughout
the unit disk $|\xi| < 1$ is:
\begin{equation}
F(a,b;c;\xi) = 1 + \frac{ab}{c}\; \left( \frac{\xi}{1!} \right) +
\frac{a(a+1)b(b+1)}{c(c+1)} \; \left(\frac{\xi^2}{2!} \right) + \cdots ,
\label{HypSeries}
\end{equation}
for $c \ne 0,-1,-2,\dots$. The function is defined elsewhere in the
complex $\xi$ plane by analytic continuation from this series expression.

A convenient way to describe the properties of $F$ for $|\xi| > 1$ is
to interchange the interior and exterior of the unit circle using the
mapping $\xi \to 1/\xi$. Because this is a particular element of the
Hypergeometric group, which maps the three singular points of
Eq.~\pref{Gauss} ($\xi = 0,1,\infty$) among themselves, it has a
simple representation on the functions $F(a,b;c;\xi)$, given
explicitly by the following identity \cite{Bateman}:
\begin{eqnarray}
\label{InvIdent}
&&\frac{F(a,b;c;\xi)}{\Gamma(c)} =\\
&&= \frac{\Gamma(b-a)}{\Gamma(b)\Gamma(c-a)}
\; (-\xi)^{-a} \, F\left(a,1-c+a;1-b+a;\xi^{-1}\right)+\nonumber \\
&&+\frac{\Gamma(a-b)}{\Gamma(a)\Gamma(c-b)}\; (-\xi)^{-b} \,
F\left(b,1-c+b;1-a+b;\xi^{-1}\right). \nonumber
\end{eqnarray}
This identity often can be usefully combined with the series expression,
Eq.~\pref{HypSeries} to obtain asymptotic forms as $|\xi| \to \infty$.

The series solutions given by eqs.~\pref{HypSeries},~\pref{InvIdent}
are calculated numerically very slowly from both sides in the vicinity of
the Alfv\'en resonance, $\xi\lsim 1$ and $\xi\ga 1$ respectively.  For
these regions we have used other representations of the Hypergeometric
functions~\cite{Bateman}.  For $\xi\lsim 1$:
\begin{eqnarray}
\label{less}
&&\frac{F(a,b;c;\xi)}{\Gamma(c)} =
\frac{\Gamma(c -a -b)}{\Gamma(c- a)\Gamma(c-b)}\times\nonumber\\ 
&&F\left(a,b;a+b-c+1;1 - \xi\right) \nonumber\\ 
&& +\frac{\Gamma(a+b-c)}{\Gamma(a)\Gamma(b)}\; (1-\xi)^{c-a-b}\times\nonumber\\
&&\times F\left(c-a,c-b;c-a-b+1;1-\xi\right)~, 
\end{eqnarray}
and for $\xi\ga 1$:
\begin{eqnarray}
\label{more}
&&\frac{F(a,b;c;\xi)}{\Gamma(c)} =
\frac{\Gamma(c -a -b)}{\Gamma(c- a)\Gamma(c-b)}\, \times\nonumber\\
&& F\left(a,a-c+1;a+b-c+1;1 - \xi^{-1}\right) \nonumber\\ 
&& +\frac{\Gamma(a+b-c)}{\Gamma(a)\Gamma(b)}\; (1-\xi)^{c-a-b}\,\xi^{a-c} \times\nonumber\\
&& F\left(c-a,1-a;c-a-b+1;1-\xi^{-1}\right)~. 
\end{eqnarray}

Another asymptotic form of the Hypergeometric function which
proves useful for our purposes is the second Watson's form for
large values of the parameters $a,b$ and finite $c$.  We use this
in its following variant \cite{Bateman}:
\begin{eqnarray}
\label{watson2} 
&&F(a+ \lambda, b- \lambda;c;\frac{1}{2} -
\frac{z}{2})= \nonumber\\
&&=\frac{\Gamma (1 - b + \lambda)\Gamma (c)}{\Gamma
(1/2)\Gamma (c-b +\lambda)}2^{a+b-1}(1 - u)^{-c +
\frac{1}{2}}\times \nonumber\\&&\times (1 + u)^{c - a - b-
\frac{1}{2}}\times \lambda^{-1/2}\left[u^{-(\lambda - b)} + e^{\pm
i\pi(c-1/2)}u^{(\lambda + a)}\right]\nonumber\\&&\times \left(1 +
O(\mid\lambda\mid^{-1})\right)~.
\end{eqnarray}
Here the argument $\xi$ is connected with the auxiliary variable
$z = 1-2\xi$ and corresponds to $u= [z + \sqrt{z^2-1}]^{-1}$ (if
$\Im \, z<0$) and to $u= [z + \sqrt{z^2-1}]$ (if $\Im \, z>0$)
with the same rule for the choice of upper (lower) sign in the
exponential factor within brackets.

The correction $O(\lambda^{-1})\ll 1$ turns out to be important for
the case $\lambda=K_{\perp}\gg 1$ which can be used for high frequency
spectrum $\omega_1\lsim N$ (although we do not present this result
here).  We find the following correction in \cite{Watson2}:
\begin{eqnarray} \label{1918} &&O= \frac{1}{2\lambda}\times \frac{R
+Te^{\pm i\pi(c-\frac{1}{2})} u^{a-b+2\lambda}}{1 + e^{\pm
i\pi(c-\frac{1}{2})}u^{a-b+2\lambda}}~,\nonumber\\&&
R=\frac{1}{2}\frac{L+ Mu^{-1}+ Nu^{-2}}{1- u^{-2}}~,\nonumber\\&&
T= \frac{1}{2}\frac{L+ Mu+ Nu^2}{1- u^{2}}~,\nonumber\\&&
L= (a+b-2c+1)^2 -a +b-\frac{1}{2}~,\nonumber\\&&
M= -2(a+b-1)(a+b-2c+1)~,\nonumber\\&&
N=(a+b-1)^2 +a -b + \frac{1}{2}~.
\end{eqnarray}

We use this expression obtaining the low frequency spectrum
(\ref{mainspectrum1}), $\omega_1(n)\ll N$, and elsewhere in the
derivation of approximate forms.

\section*{Appendix B. Asymptotic Forms}
The identities of the preceding Appendix~A can be used to derive the
asymptotic forms in the limit of large $\xi$ for the solutions
$Y_1$ and $Y_2$ discussed in the main text. For these purposes we
use the following more convenient basis of solutions to
Eq.~\pref{Gauss}:
\begin{eqnarray}
\label{xilarge1}
    &&Y_3= \xi^{-b} F\left( b+1-c,b;a+b-c+1;1-
    \frac{1}{\xi} \right)~,\nonumber\\
    &&Y_4=\xi^{-b}\left( \frac{1}{\xi}-1 \right)^{c-a-b}\times\nonumber\\
&&  \times  F\left( c-a,1-a;c-a-b+1;1-\frac{1}{\xi} \right)~ .
\end{eqnarray}
The expansion of our original solutions, $Y_1$ and $Y_2$, in terms
of these new ones has the form $Y_1 = N_1 \, Y_3 + N_2 \, Y_4$ and
$Y_2 = N_1' \, Y_3 + N_2' \, Y_4$, with coefficients $N_{1,2}$ and
$N_{1,2}'$ given by
\begin{eqnarray}
&& N_1=\frac{\Gamma(c)\Gamma(c-a-b)}{\Gamma(c-a)\Gamma(c-b)}\,,\nonumber\\
&& N_2=\frac{\Gamma(c)\Gamma(a+b-c)}{\Gamma(a)\Gamma(b)}\,,\nonumber\\
&& N_1'=\frac{\Gamma(2-c)\Gamma(c-a-b)}{\Gamma(1-a)\Gamma(1-b)}\,,\nonumber\\
&& N_2'=\frac{\Gamma(2-c)\Gamma(a+b-c)}{\Gamma(a+1-c)\Gamma(b+1-c)}\,.
\label{Ncoefficients}
\end{eqnarray}

The large-$\xi$ limit is now easily obtained using the result
\[
F(a,b;c;1)=\frac{\Gamma(c)\Gamma(c-a-b)}{\Gamma(c-a)\Gamma(c-b)}~,
\]
which is finite provided $\Re~(c-a-b)>0$ and $c \neq 0,-1,-2,...$
\cite{Bateman}. Notice that these conditions are satisfied for the
choices of $a,b$ and $c$ obtained in the main text. Applying these
expressions to $Y_3$ and $Y_4$ then gives, in the limit $\xi\to
\infty$:
\begin{eqnarray}
    Y_3 &\to& \xi^{-b} \frac{\Gamma(1+a+b-c)\Gamma(a-b)}{\Gamma(a)
    \Gamma(a+1-c)}~, \nonumber \\
    Y_4 &\to& \xi^{-b}(-1)^{c-a-b} \frac{\Gamma(1-a-b+c)
    \Gamma(a-b)}{\Gamma(1-b)\Gamma(c-b)}~.
\end{eqnarray}
Using the previously-given expression, Eq.~\pref{Ncoefficients},
for the coefficients $a,b$ and $c$, the limiting form for original
solutions $Y_{1,2}(\xi)$ finally become:
\begin{eqnarray}
&&Y_1\to
\xi^{-b}\frac{\Gamma(c)\Gamma(a-b)}{\Gamma(a)\Gamma(c-b)}l_1\,,\nonumber\\
&& Y_2\to
\xi^{-b}\frac{\Gamma(2-c)\Gamma(a-b)}{\Gamma(1-b)\Gamma(1+a-c)}l_2~,
\label{YlimitB} 
\end{eqnarray}
with coefficients $l_{1,2}$ given by the elementary functions
\begin{eqnarray}
&& l_1= \frac{\sin \pi(c-a)- (-)^{c-a-b}\sin \pi
    b}{\sin \pi(c-a-b)}\,,\nonumber\\
&& l_2= \frac{\sin \pi a- (-)^{c-a-b}\sin
    \pi (c-b)}{\sin \pi(c-a-b)}~.
\label{elementary}
\end{eqnarray}
Notice that these quantities satisfy the relation $l_2/l_1=
-e^{i\pi c}$.


\label{lastpage}
\end{document}